\documentclass{aa}  

\usepackage{graphicx}
\usepackage{txfonts}
\usepackage{subcaption}
\usepackage{hyperref}
\usepackage{amsmath}
\usepackage{xcolor}
\usepackage{placeins}

\emergencystretch=\maxdimen     
\begin{document} 

   \title{Deep radio interferometric search for decametre radio emission from the exoplanet Tau Bo\"{o}tis b}

   \author{C.M. Cordun
          \inst{1,2}
          \and
          H.K. Vedantham\inst{1,2}
          \and
          M.A. Brentjens \inst{1}
          \and 
          F.F.S. van der Tak \inst{3,2}
          }

   \institute{ASTRON, Netherlands Institute for Radio Astronomy, Oude Hoogeveensedijk 4, Dwingeloo, 7991PD, The Netherlands\\
              \email{cordun@astron.nl}
         \and
             Kapteyn Astronomical Institute, University of Groningen, PO Box 800, 9700 AV, Groningen, The Netherlands
        \and 
            SRON Netherlands Institute for Space Research, Landleven 12, NL-9747AD Groningen, The Netherlands
             }

   \date{Received November 4, 2024; Accepted December 16, 2024}

\abstract
   {Detection of electron cyclotron maser (ECM) emission from exoplanets in the 10-40~MHz radio band is likely the only way to measure an exoplanet's magnetic field directly. However, no definitive detection of exoplanetary ECM emission has been made to date. A detection of the hot Jupiter Tau Bo\"otis b was reported but with an observing mode that is not immune to confusion from off-axis interference, making the detection tentative.}
   {We searched for radio emissions from Tau Boötis b using the Low Frequency Array (LOFAR) in interferometric mode, which employs direction-of-arrival information to discriminate genuine signals from interference. Our aim was to confirm the previous tentative detection or establish an upper limit in the case of a non-detection.}
   {We conducted observations using LOFAR’s Low Band Antenna in interferometric mode, which totalled 64\, hours spread over 8\, nights. We created a custom data-processing pipeline to mitigate common challenges in decametric radio astronomy, including radio frequency interference, ionospheric distortions, and sidelobe noise from nearby bright radio sources. We used this pipeline to image the field around Tau Bo\"otis b, searching for both quiescent and bursting emission from the direction of Tau Bo\"otis b.}
   {Despite the high sensitivity of the interferometric observations and extensive data processing, no significant emission was detected from Tau Boötis b in Stokes V. We establish an upper limit of 2 sigma at 24~mJy for any continuous emission from the exoplanet. The previous tentative detection of 400~mJy is thus not confirmed by the interferometric observations.}
   {The previous tentative detection is unlikely to be a bona fide astrophysical signal. Our upper limit is lower than the flux density predicted by scaling laws, meaning either the scaling laws need to be revised or the emission from this particular system is beamed away from Earth.}
   

   \maketitle

%

\section{Introduction}
\nolinenumbers
Jupiter is the brightest object in the sky at frequencies below 40~MHz with the exception of solar bursts. This large flux arises primarily due to its strong magnetic field and the large number of electrons trapped in its magnetosphere, originating from Io's volcanic activity \citep{dungey1961steady, zarka2001magnetically}. These electrons are accelerated by a breakdown of the co-rotation of Jupiter's magnetic field and its electrodynamic interaction with its large moons, notably Io \citep{zarka1992auroral, zarka1998auroral}. Additionally, there is a solar-wind-mediated component to Jupiter's decametric emission. This is particularly relevant for understanding Jupiter-like exoplanets, as such planets may lack moons capable of creating a plasma torus like Io does. However, they will still interact with their host star's stellar wind, generating radio flux density \citep{2007P&SS...55..598Z}. The resulting radio emissions, driven by the electron cyclotron maser (ECM) process \citep{dulk1985radio}, are expected to be detectable only at low frequencies \citep[$\nu\lesssim 40 {\rm~MHz}$;][]{griessmeier2007predicting}.

Because ECM emission is beamed and originates at the poles, it appears to be periodically modulated to an observer.  In the co-rotation breakdown scenario, the emitting electrons are present at all magnetic longitudes, yielding a periodic modulation at the planet's rotational period \citep{higgins1996new}. In contrast, the emitting electrons are confined to the flux tube connecting Jupiter and Io \citep{bigg1964influence} in the planet--satellite interaction scenario. In this case, the ECM emission period matches the satellite's orbital period \citep{ip2004star, lanza2012star}.

The maximum frequency ($f_c^{max}$) of ECM emission is directly proportional to the planet's polar magnetic field strength ($B$) at the surface: 
\begin{equation}
    f_c^{max}[MHz] = 2.8 B [G]
    \label{eq:1}
.\end{equation}

\noindent Detection of this peak frequency is one of the most promising techniques for directly measuring an exoplanet's magnetic field strength, as it provides the most model-independent method compared to other approaches. For example, Jovian ECM emission only appears below 40~MHz \citep{bagenal2017magnetospheric} due to its polar surface magnetic field strength of around 9~gauss. Additionally, ECM emission is close to 100\% circularly polarised \citep{zarka2007plasma}, making it easier to detect against a background of extragalactic radio sources that are expected to be linearly polarised or unpolarised.

To date, magnetised Solar System planets are the only planets with confirmed ECM detections \citep{dulk1985radio}. Efforts to detect radio emission from exoplanets have primarily targeted hot Jupiters because of their proximity to their host stars and predicted strong magnetic fields, which should result in high radio flux densities \citep{zarka1997ground,farrell1999possibility, zarka2007plasma}. Extensive searches have only yielded non-detections \citep{yantis1977search,winglee1986search,bastian2000search,lazio2007magnetospheric,smith2009secondary,lazio2009blind, hallinan2012looking,lynch2017search, lenc2018all, de2020radio} and a few tentative signals \citep{lecavelier2013hint,sirothia2014search,vasylieva2015pulsars,turner2021search}. Recent flux density predictions for these planets based on a simple scaling of Solar System observations are on the order of hundreds of millijanskys \citep{joseph2004radiometric,griessmeier2005influence, griessmeier2007exoplanetary, griessmeier2007predicting, zarka2007plasma, reiners2010magnetic, griessmeier2017search,lynch2018detectability, zarka2018jupiter}, but if more conservative saturation scenarios are considered \citep{turnpenney2020magnetohydrodynamic}, then these predictions fall to below 100~mJY for the brightest hot Jupiters.

The magnetic properties of hot Jupiters remain largely unknown. While these planets have been observed in other wavelengths such as optical and infrared \citep{seager2005dayside, fortney2006influence, cowan2007hot, parmentier2016transitions}, radio observations are crucial for characterising their magnetic properties and space weather conditions. The reasons for non-detections are still unclear, with potential explanations including insufficient instrument sensitivity \citep{zarka2015magnetospheric}, observing frequencies not low enough \citep{griessmeier2005influence}, Earth positioned outside the emission beam \citep{hess2011modeling}, or incorrect scaling laws that do not properly predict their flux densities \citep{weber2017cyclotron,weber2017expanded,weber2018supermassive,kavanagh2019moves}.

Tau Boötis b, a gas giant detected via the radial velocity method \citep{duquennoy1991multiplicity, butler1997three}, has garnered significant attention. This planet is larger and more massive than Jupiter, suggesting a stronger magnetic field \citep{reiners2010magnetic}. Its proximity to its host star and Earth makes it the ideal candidate for radio emission searches, with predictions indicating it to be the brightest known radio exoplanet \citep{griessmeier2007predicting, lynch2018detectability}. A tentative detection using the Low Frequency Array \citep[LOFAR; ][]{VanHaarlem2013} was reported by \citet{turner2021search}, but follow-up observations did not re-detect this signal \citep{turner2023follow}.

LOFAR can observe in two modes: beam-formed and interferometric \citep{VanHaarlem2013}. The beam-formed mode uses multiple beams (typically four: one on the source and three controls on an `empty' part of the sky) and only records the radiometric power as a function of time, frequency, and polarisation. The control-beam or off-beam data are compared with the on-beam data to identify signals. However, this mode lacks direction-of-arrival information, making it susceptible to human-generated interference, that is, radio frequency interference (RFI), and other spurious off-axis interference. In contrast, the interferometric mode records the complex spatial coherence of the electric field, which preserves direction-of-arrival information. The mode also allows us to correct instrumental and ionospheric effects offline. The beam-formed mode's lack of spatial localisation therefore necessitates confirmation of any detection via the interferometric mode.

This study presents a follow-up observation of \citet{turner2021search} from an interferometric perspective. Previous decametre-wavelength studies of exoplanets have relied on the beam-formed mode \citep{turner2021search, turner2023follow}, which, as noted, has significant localisation drawbacks, especially in the RFI-dominated bands below 25~MHz. The interferometric mode aims to resolve these ambiguities and create precise sky images to localise any excess Stokes V emission. 

For this study, Tau Boötis b was observed using LOFAR and its Low Band Antenna (LBA), between 15 and 40~MHz in the interferometric mode. Observations were conducted in 2020 during the solar minimum, for 56 hours, centred around the planet’s maximum elevation. These interferometric observations were obtained at the same time as the beam-formed observations from \cite{turner2023follow}. Only the core stations were used, limiting the longest baseline to 1 km and resulting in a resolution of half a degree. LOFAR allows imaging in all polarisations, which is essential since Stokes I is confusion-limited at these low resolutions. In contrast, the sky is almost empty in Stokes V, leaving exoplanets as one of the few detectable sources.

Observing exoplanets at frequencies below 40~MHz presents several challenges, including RFI, ionospheric phase corruption, and dynamic range issues due to the presence of bright radio sources. Below 25~MHz, nearby electronics such as electric fences and power sources dominate the spectrum \citep{Offringa2013}, making up to half of observations unusable. Earth's ionosphere introduces a time-variable dispersive delay whose higher-order terms must also be calibrated for frequencies below 40\,~MHz \citep{de2018effect}. Additionally, due to the large primary beams at low frequencies, bright radio sources like Cassiopeia A, Cygnus A, Taurus A, Virgo A, and Hercules A (the A-team sources) create systematic sidelobe noise \citep{de2020cassiopeia}. Specifically, Virgo A's proximity to Tau Boötis b (approximately 10 degrees) complicates calibration. The goal of this study is to obtain the lowest-frequency images to date, thus expanding the parameter space for decametre science, such as galaxies, the interstellar medium, and other phenomena.

The article is structured as follows. Section \ref{sec:obs} describes the observations in detail, while Sect. \ref{sec:data_proc} outlines the data-processing pipeline developed for core station observations at decametre wavelengths. Section \ref{sec:results} presents the results, including calibrated images, dynamic spectra, and noise analysis. Section \ref{sec:discussion} discusses the upper limits, places the observations in a broader context, and comments on the previous tentative detection. Finally, concluding remarks are provided in Sect. \ref{sec:conclusion}.


\section{Observations}\label{sec:obs}
All observations presented here were acquired using the LOFAR \citep[][]{VanHaarlem2013} LBA system simultaneously in beam-formed and interferometric modes. They were conducted as a follow-up to the initial tentative detection reported by \citet{turner2021search} in beam-formed data. Due to this dual-mode operation, the system could only use the core stations. The beam-formed observations have already been analysed and published by \citet{turner2023follow}. This paper focuses on analysing the interferometric data.

\begin{figure}[htb]
    \centering
    \includegraphics[width = 0.9\linewidth]{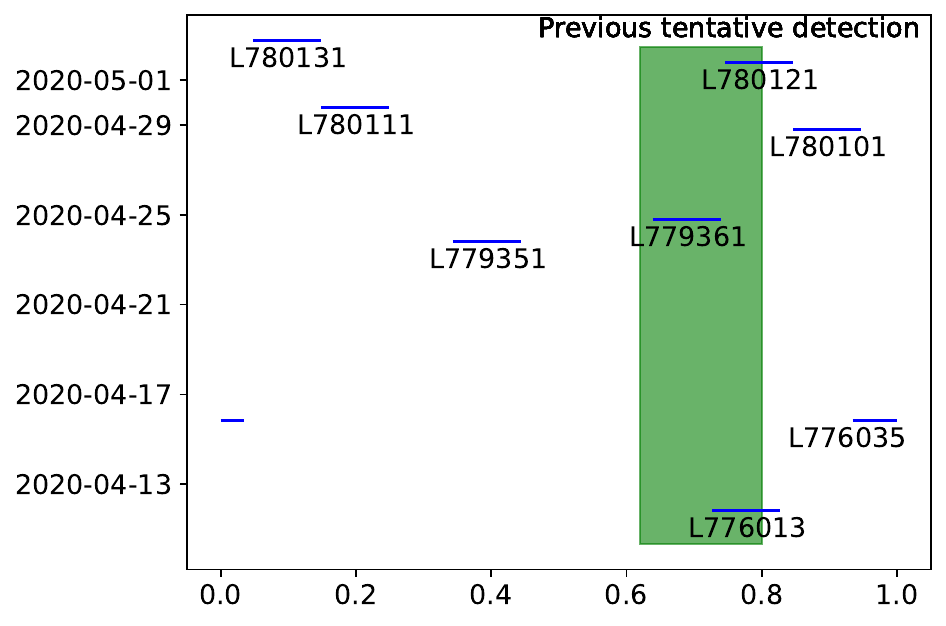}
    \caption{Planet's phase during the LOFAR observation. The area corresponding to the previous detection is marked with a green rectangle. Similar to \citet{turner2021search}, we calculate the orbital phases relative to the periastron, where the planet's orbital period is 3.31 days and the periastron time is 2446957.81JD \citep{wang2011eccentricity}. The blue bar on the lower left is a continuation of observation L776035.}
    \label{fig:phase-planet}
\end{figure}

The dataset contains eight observations, each spanning 8 hours, from 14.8 to 39.8~MHz. These observations used 21 out of the 24 LOFAR core stations in the \texttt{LBA-Outer} mode \citep{VanHaarlem2013}, and the target and the calibrator were observed simultaneously. Detailed configuration parameters are provided in \autoref{tab:array-details}. The observational schedule was arranged to coincide with the planetary phases of the previous tentative detection and where radiation was predicted by \citet[see our Fig. \ref{fig:phase-planet}]{ashtari2022detecting}. The epochs had a cadence of a few days and were centred around the target's maximum elevation. Observation details and corresponding planetary phases of each dataset are listed in \autoref{tab:obs-details}.

\begin{table}[!ht]
\centering
\caption{Setup of the LOFAR array during the observations.}
\label{tab:array-details}
\begin{tabular}{c|c}
\hline\hline
Parameter              & Value \\
\hline
Instrument             & LOFAR-LBA     \\
Array setup            & LOFAR Core  \\
Number of stations     & 21          \\
Antennas configuration & LBA Outer \\
Minimum frequency      & 14.8~MHz    \\
Maximum frequency      & 39.8~MHz    \\
Sub-bands per dataset               & 122         \\
Channels per sub-band   & 64          \\
Frequency resolution   & 3.2 kHz   \\
Time resolution        & 1 s         \\
Resolution at 14.8~MHz & 0.46 deg    \\
Resolution at 38.8~MHz & 0.17 deg   \\
Calibrator             & 3C295 \\
\hline
\end{tabular}
\end{table}

\begin{table}[!ht]
\centering
\caption{Details of the 8-hour observations.$\rm ^{(a)}$}
\label{tab:obs-details}
\begin{tabular}{ccc}
\hline\hline
LOFAR SAS ID & \begin{tabular}[c]{@{}c@{}}Start Date and \\ Time (UTC)\end{tabular} & Phase     \\ \hline
L776013      & 2020-04-14 20:00                                                     & 0.73–0.83 \\
L776035$\rm ^{(b)}$      & 2020-04-15 20:00                                                     & 0.93–0.03 \\
L779351      & 2020-04-23 19:30                                                     & 0.34–0.44 \\
L779361      & 2020-04-24 19:00                                                     & 0.64–0.74 \\
L780101      & 2020-04-28 19:00                                                     & 0.85–0.95 \\
L780111      & 2020-04-29 19:00                                                     & 0.15–0.25 \\
L780121      & 2020-05-01 18:30                                                     & 0.75–0.85 \\
L780131      & 2020-05-02 18:30                                                     & 0.05–0.15 \\
\hline
\end{tabular}
\tablefoot{$\rm ^{(a)}$ Similar to \citet{turner2021search}, we calculate the orbital phases relative to the periastron, where the planet's orbital period is 3.31 days and the periastron time is 2446957.81JD \citep{wang2011eccentricity}.}
\tablefoot{$\rm ^{(b)}$ This dataset could not be properly calibrated due to intense ionospheric phase and amplitude scintillations.}

\end{table}

The calibrator used for this analysis was 3C295, which in retrospect, was a suboptimal choice due to its spectral turnover: its flux density drops below 30 Jy at frequencies around 15~MHz \citep{scaife2012broad}. However, this challenge was fortuitously mitigated by low ionospheric activity during the observations, which can be seen in the lack of rapid phase-wraps in the calibrator's phase solutions for the longest baseline presented in \autoref{sec:phasesol}. The observations were mostly scheduled at night when the levels of RFI reflected off the ionosphere are minimal. Additionally, the solar minimum ensured that the ionospheric phase errors were modest, facilitating calibration. While using only the core stations limits angular resolution and sensitivity, it simplifies the calibration process. These factors ensure that the data can be calibrated to the lowest LOFAR frequencies, even with 3C295 as the calibrator.

\section{Data processing}\label{sec:data_proc}

We calibrated the data using a strategy similar to that of the LOFAR Initial Calibration (LINC) Pipeline \citep{de2019systematic} and the Library for Low Frequencies \citep{de2020reaching}. However, both pipelines are ineffective for data with only core stations and at frequencies below 40~MHz, necessitating the development of a custom pipeline available on GitHub\footnote{\footnotesize{\url{https://github.com/cristina-low/lofar-low-pipelines.git}}}.

\begin{figure}[!htb]
    \centering
    \includegraphics[width = 0.9\linewidth]{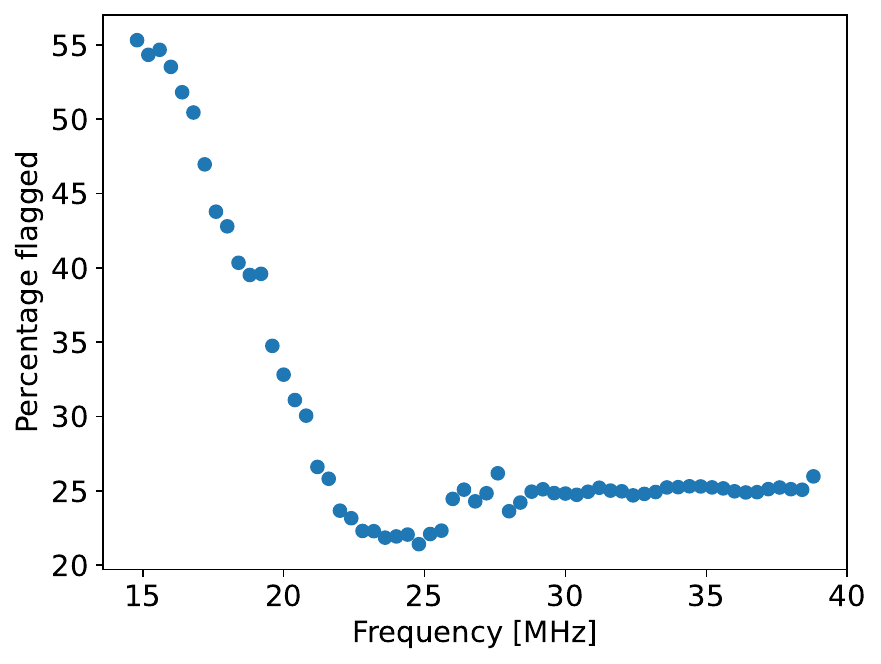}
    \caption{Percentage of flagged data due to RFI as a function of frequency.}
    \label{fig:flags}
\end{figure}

The pipeline starts by flagging RFI with the default \texttt{AOFlagger} strategy \citep{offringa2010aoflagger}, which identifies narrow, bright radio emission and removes it from the visibilities. Next, we removed (de-mixed) five bright radio sources: Cassiopeia A, Cygnus A, Taurus A, Virgo A, and Hercules A, using models from \citet{de2020cassiopeia}. This step is challenging due to Virgo A's proximity to our phase centre (10 degrees) and the relative inefficiency of de-mixing in the presence of only short core-core baselines. To mitigate Virgo A's artefacts in our field of view, we used larger intervals for de-mixing than is standard practice. We used a time resolution of 30 seconds and a frequency resolution of 400 kHz. A subsequent flagging step on full bandwidth removed broadband interference, resulting in 20\% to 55\% flagged data, depending on the frequency. The average flagging rates as a function of frequency are shown in Fig. \ref{fig:flags}.

\begin{figure*}[!t]
    \centering
    \begin{subfigure}[t]{0.47\textwidth}
        \centering
        \includegraphics[width = 0.98\linewidth]{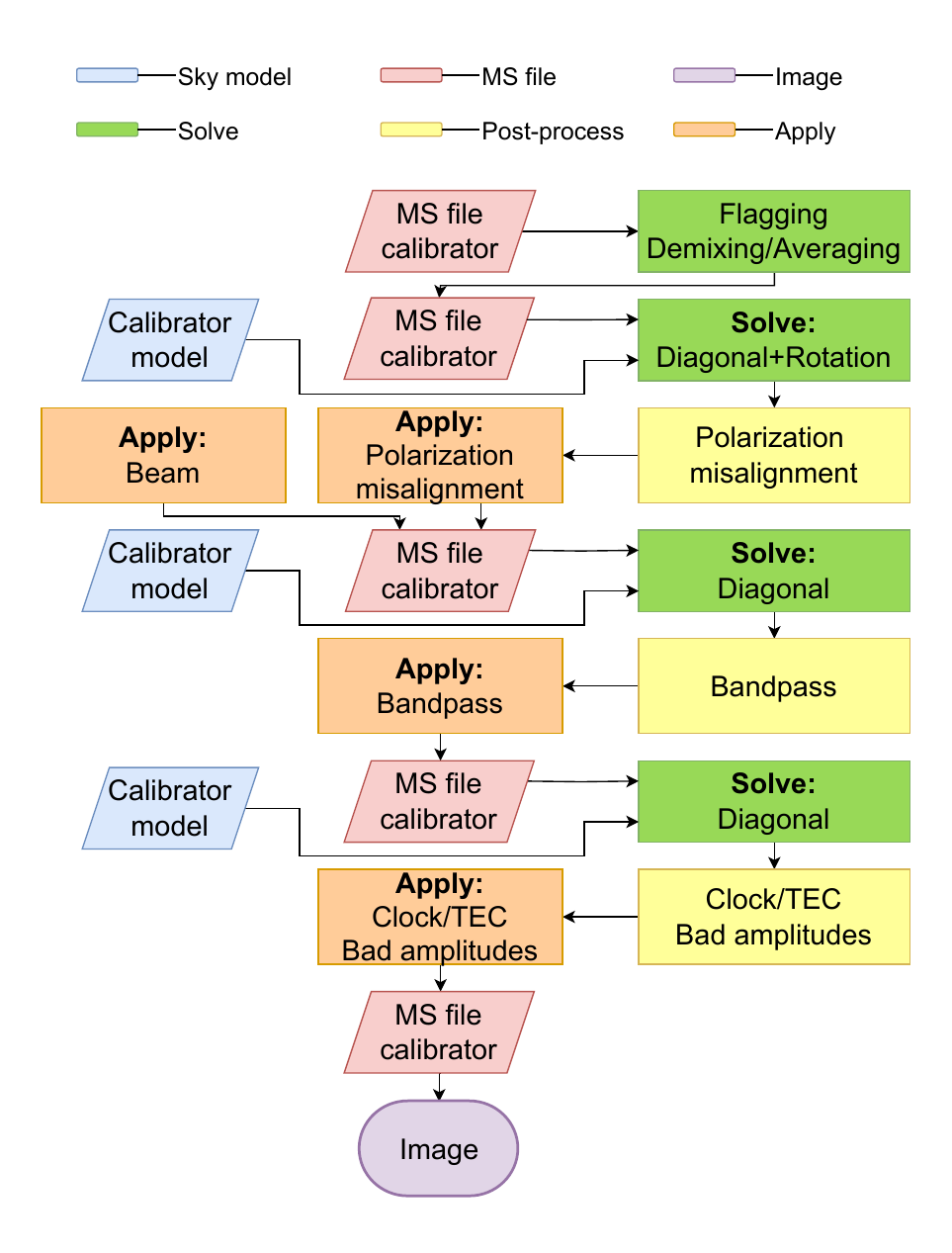}
    \end{subfigure}%
    ~ 
    \begin{subfigure}[t]{0.47\textwidth}
        \centering
        \includegraphics[width = 0.98\linewidth]{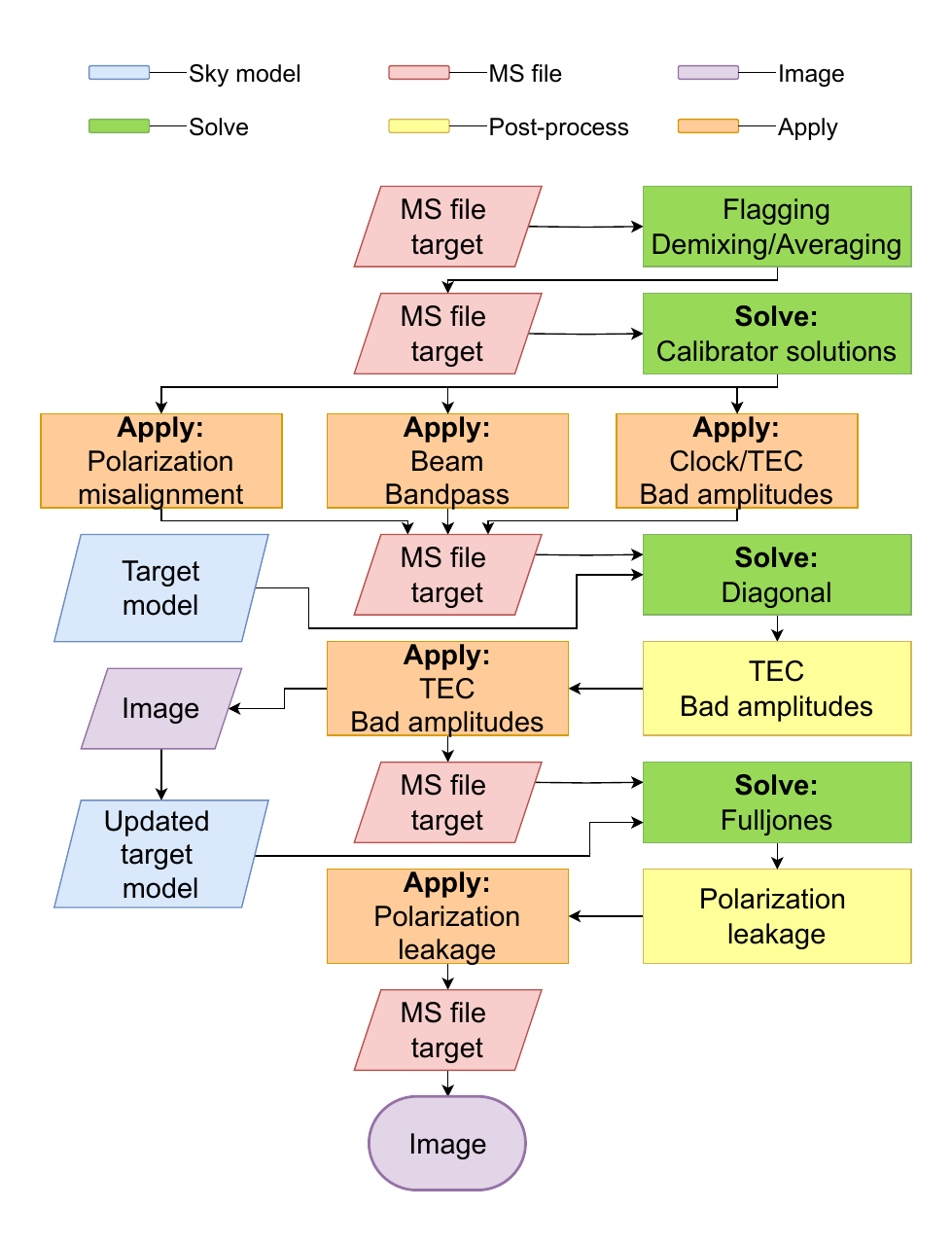}
    \end{subfigure}
    \caption{Summary of the calibration strategy used to process the calibrator field (left) and the target field (right) for data below 40~MHz from core stations only.}
    \label{fig:pipelines}
\end{figure*}

Next, we used 3C295 to correct for polarisation misalignment, beam pattern effects, and bandpass using the Default Preprocessing Pipeline (\texttt{DP3}) of LOFAR and the LOFAR solutions tool (\texttt{LoSoTo}), using a procedure similar to \citep{de2019systematic}. We did not correct for differential Faraday rotation as its effects are negligible because the observations involved only core stations and were recorded during the solar minimum.

The next calibration step involves separating the stations' clock delays and ionospheric delays on the calibrator. We could not use the associated \texttt{LoSoTo} function as its solutions did not converge to meaningful values, likely due to the low frequency of observation demanding the use of higher-order terms in the expression for dispersive delay. We developed a script to calculate the delays between the stations and the first- and third-order ionospheric effects \citep{mangla2023exploring}. This script exploits the fact that the core station delays are less than 10 ns (they are on a common clock) and that the ionosphere is calm due to the solar minimum, resulting in no $2\pi$ phase wraps.

The last step consists of calculating amplitude solutions on the calibrator and flagging the outliers resulting from RFI. After applying all these solutions to the calibrator field, we assessed the quality of the calibration solutions by imaging the calibrator with \texttt{WSClean} \citep{offringa2014wsclean}. A summary of the calibrator pipeline is presented in Fig. \ref{fig:pipelines} (left). For completeness, we present the calibrator images used to assess the quality of the calibration in \autoref{sec:imagesCal}.

We then applied the calibrator's solutions to the target field, including corrections for polarisation misalignment, primary beam, bandpass, and clock. Even though it is not standard practice to apply the ionospheric phase corrections in the direction of the calibrator to the target data, we made an exception due to the small angular offset between the two objects (less than 15 degrees), and calm conditions. The amplitude flags are also applied since the target and calibrator data were collected simultaneously and are subject to similar RFI.

We performed a self-calibration step on the target field to mitigate residual ionospheric phase errors and to improve the target field model. With three in-field sources of approximately 15\,Jy flux density, we chose a frequency interval of 2~MHz and a time interval of 15 minutes to achieve a signal-to-noise ratio per visibility of around 3 at 15~MHz. The 15-minute integration time preserves data coherence due to calm ionospheric conditions, resulting in phase variations of less than 1 radian per baseline over this interval. The phase solutions for the longest baseline are shown in Appendix   \ref{sec:phasetarget}.

Any remaining corruption due to RFI was removed from the amplitude solutions. The calibrated dataset was imaged with \texttt{WSClean}, and the target model was updated. Self-calibration is essential to correct residual phase differences between stations, which arise because the calibrator and target occupy different positions in the sky. Initially, the target field model includes only the brightest sources; the model is subsequently refined for the following calibration steps as fainter sources emerge in the improved images.

The resulting Stokes V images have a 10\% leakage from Stokes I. This leakage arises because the dipoles are not perfectly perpendicular for a non-zenith source, causing a common-mode signal between the X and Y hands that can manifest as Stokes-V flux if there are any residual instrumental phase offsets between the X and Y hands of the array. This is expected because thus far only diagonal Jones matrix terms were applied to the data. Thus, we performed an additional polarisation calibration step, calculating complete Jones matrices for the entire time interval for every 2.4~MHz bandwidth. We applied only the anti-diagonal terms, setting the diagonal terms to unity. This reduced the leakage from 10\% to 1\%. The fully calibrated target dataset was imaged again with \texttt{WSClean}. A summary of the target pipeline is presented in Fig. \ref{fig:pipelines} (right).

\section{Results}\label{sec:results}

\begin{figure*}[h!t]
    \centering
    \begin{subfigure}[t]{0.43\textwidth}
        \centering
        \includegraphics[width = 1\linewidth]{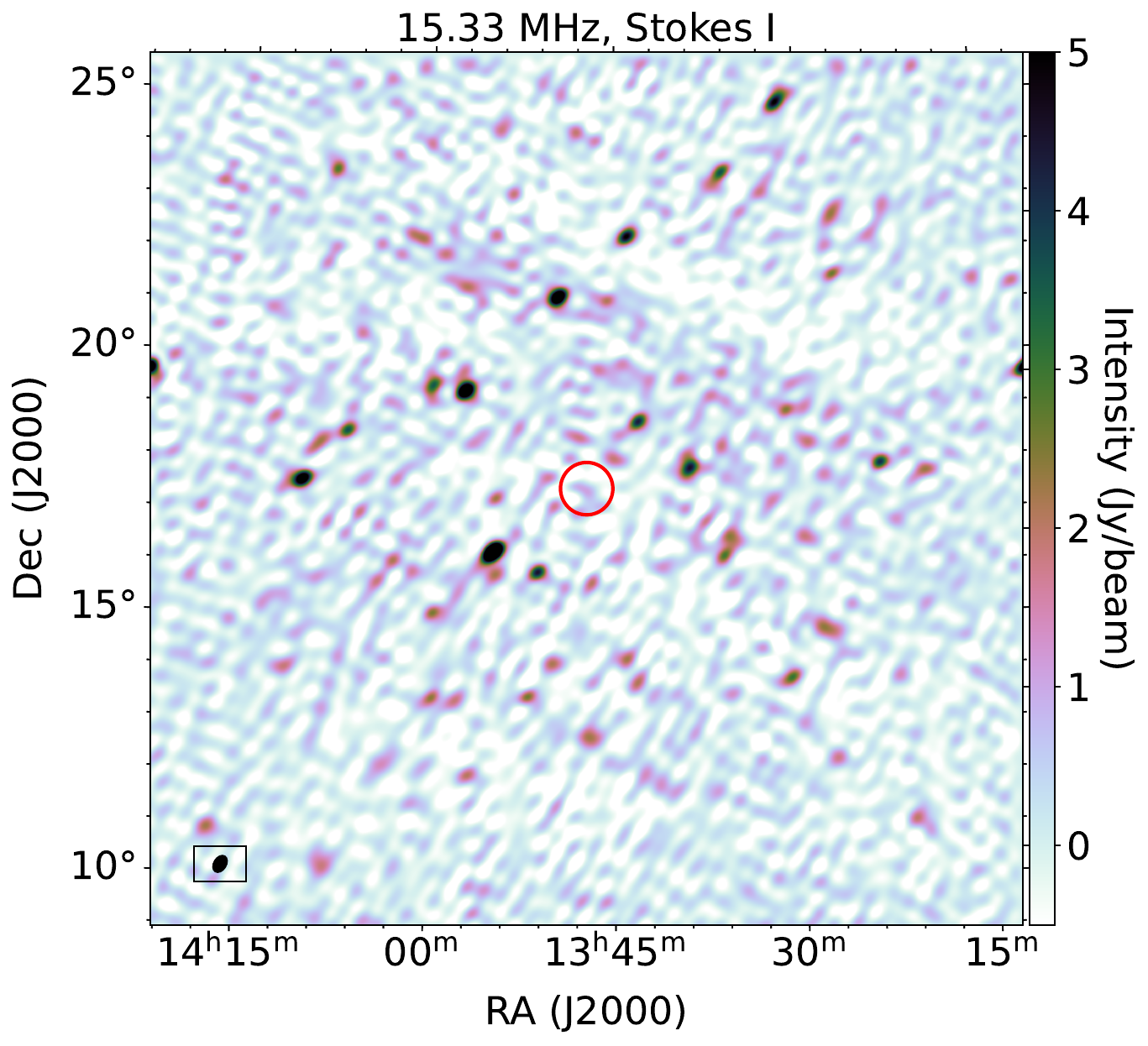}
    \end{subfigure}%
    ~ 
    \begin{subfigure}[t]{0.43\textwidth}
        \centering
        \includegraphics[width = 1\linewidth]{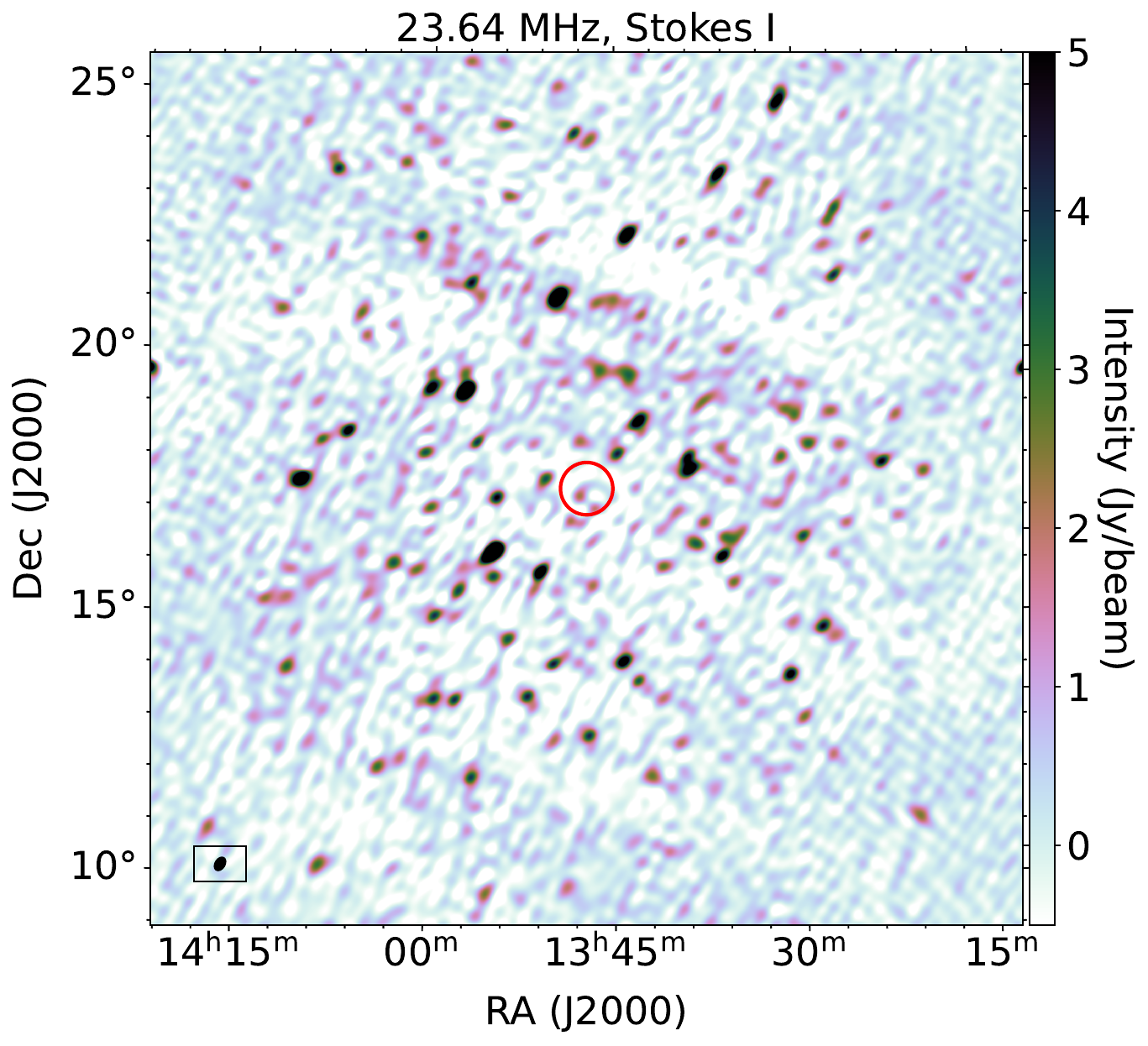}
    \end{subfigure}

    \begin{subfigure}[t]{0.43\textwidth}
        \centering
        \includegraphics[width = 1\linewidth]{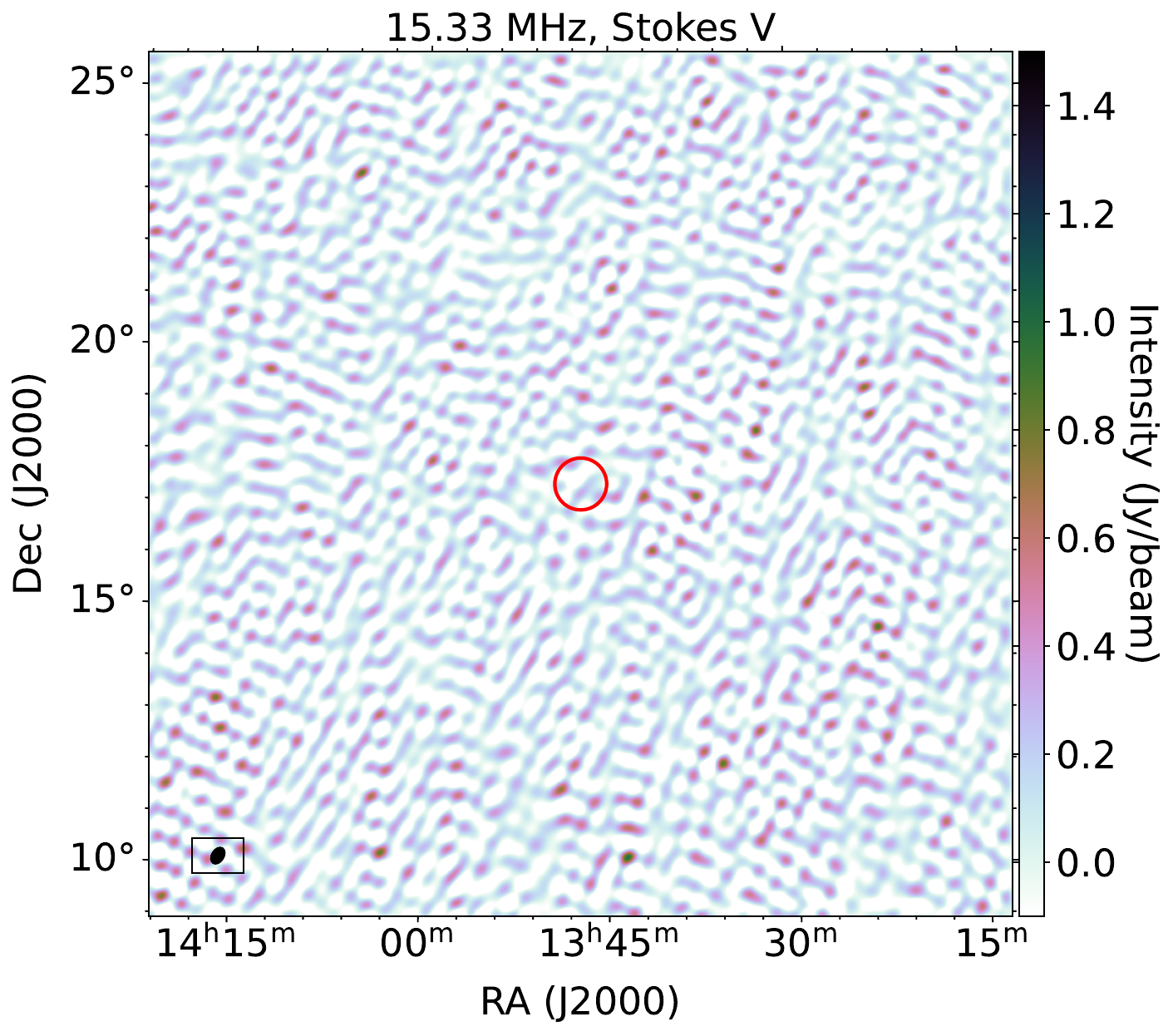}
    \end{subfigure}%
    ~ 
    \begin{subfigure}[t]{0.43\textwidth}
        \centering
        \includegraphics[width = 1\linewidth]{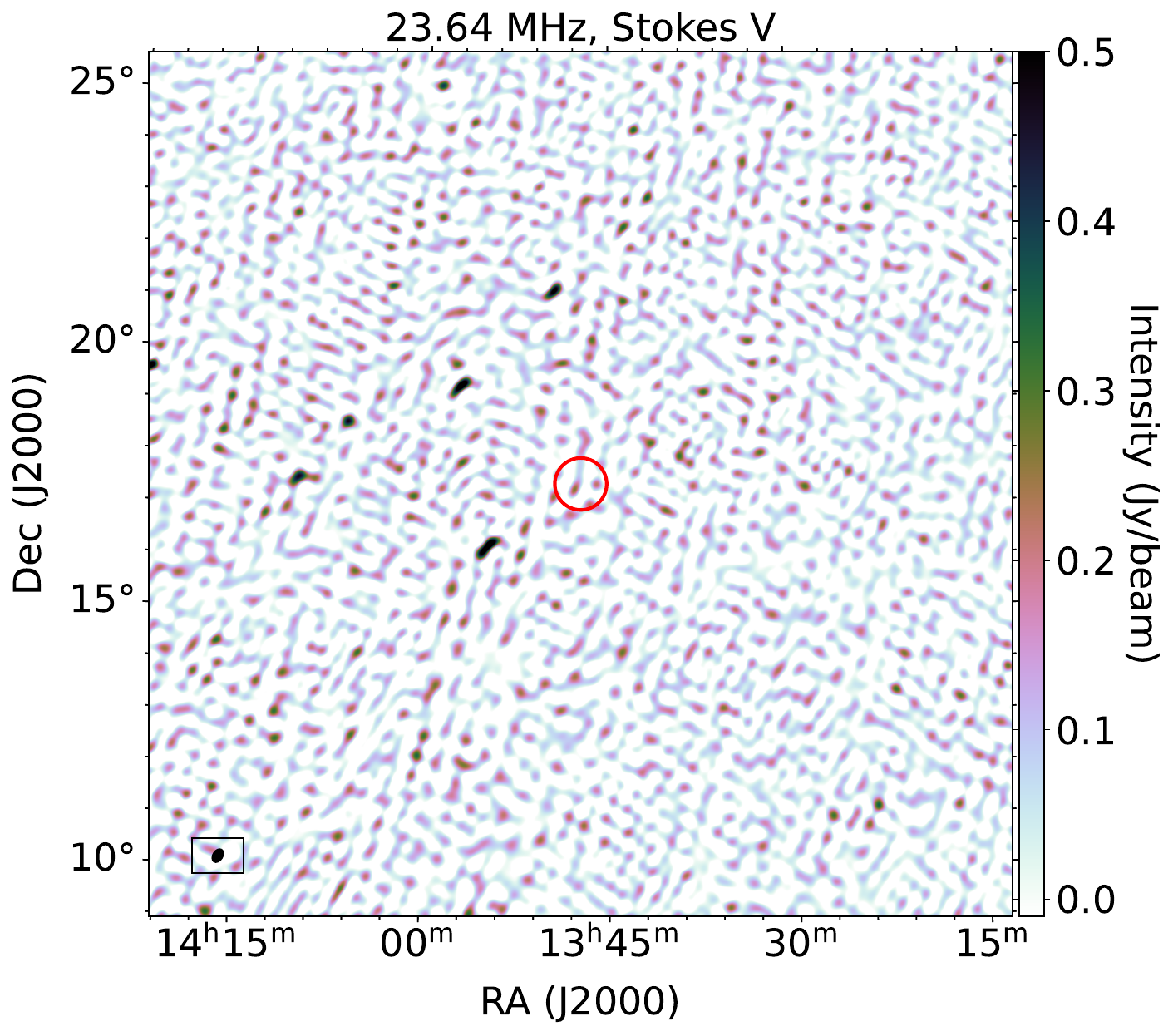}
    \end{subfigure}
    \caption{Images at 15~MHz and 24~MHz in different polarisations for an integration time of 56 hours and a bandwidth of 1.2~MHz. The planet should be at the centre of the red circle. The beam shape is in the lower-left corner.}
    \label{fig:images_freq}
\end{figure*}

\begin{figure*}[h!t]
    \centering
    \begin{subfigure}[t]{0.43\textwidth}
        \centering
        \includegraphics[width = 1\linewidth]{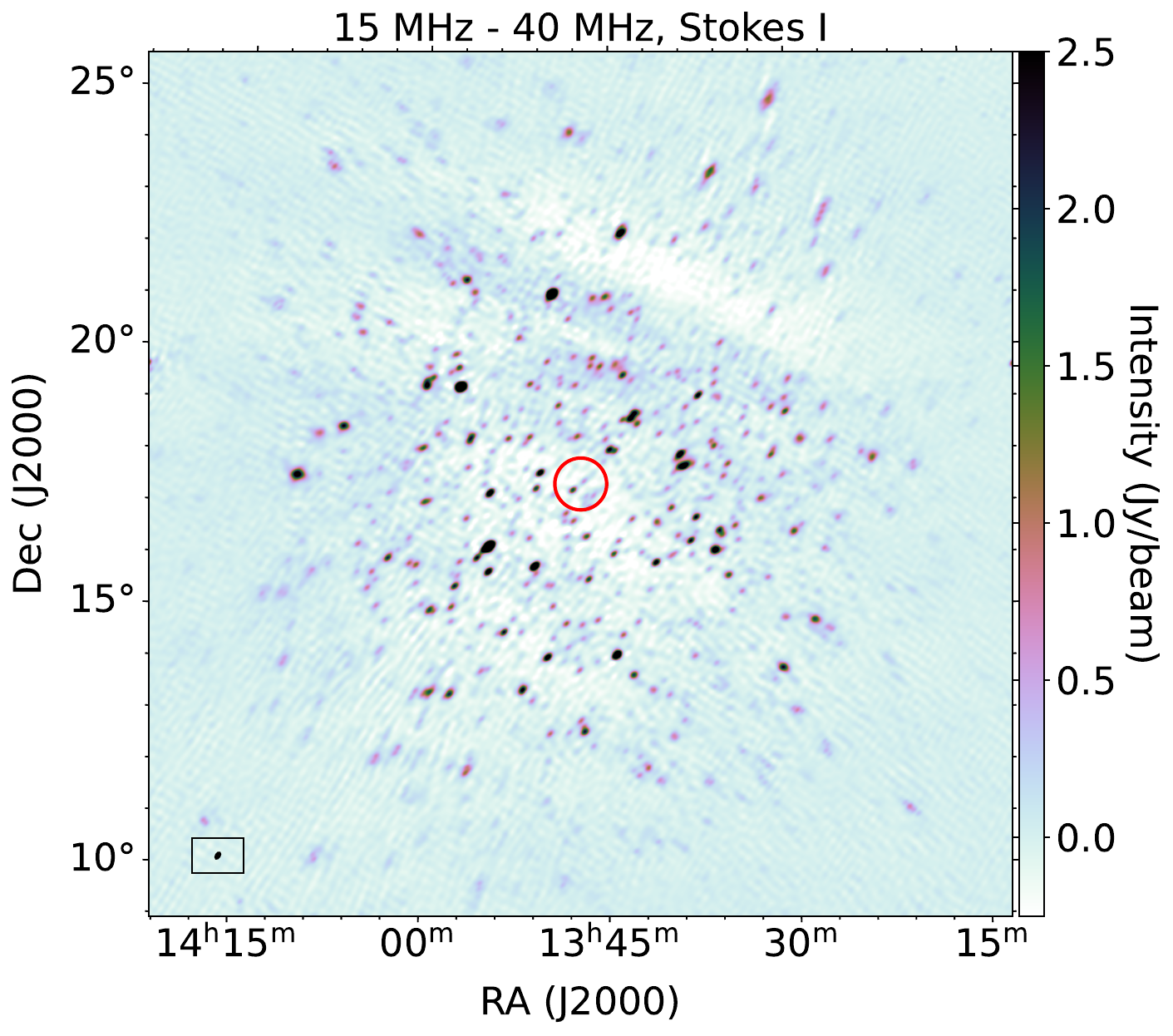}
    \end{subfigure}
    \begin{subfigure}[t]{0.43\textwidth}
        \centering
        \includegraphics[width = 1\linewidth]{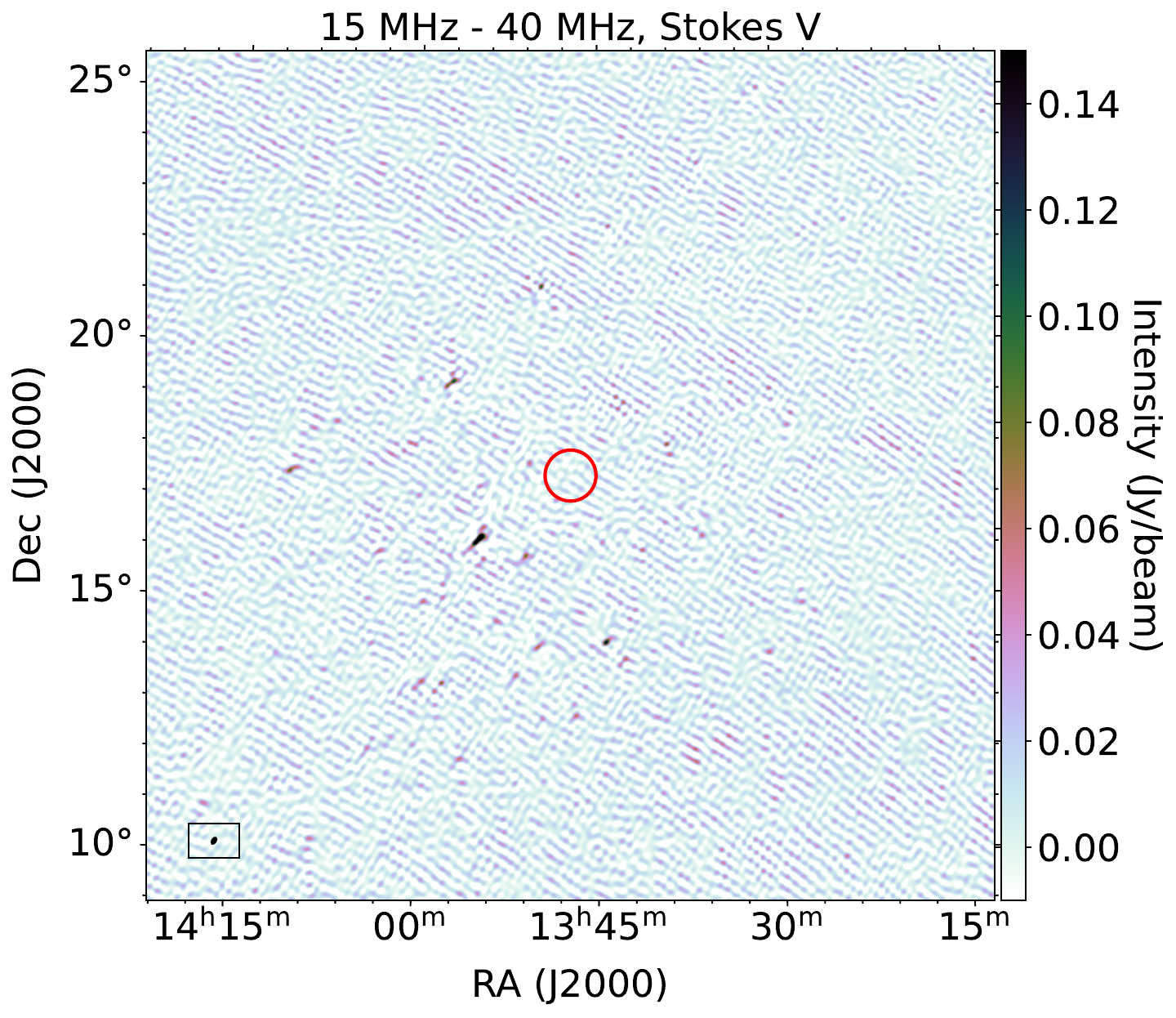}
    \end{subfigure}
    \caption{Images in different polarisations for an integration time of 56 hours and a bandwidth of 35~MHz. The planet should be at the centre of the red circle. The beam shape is in the lower-left corner.}
    \label{fig:images_BW}
\end{figure*}

We created images at 1.2~MHz intervals to search for narrow-band, faint emission from the exoplanet in Stokes I and Stokes V. Figure \ref{fig:images_freq} shows the 15~MHz images, which represent the lowest-frequency radio images at sub-degree resolution ever obtained and the 24~MHz images; they are close to the frequency of the tentative detection reported by \cite{turner2021search}. We also checked for faint broadband emission using Fig. \ref{fig:images_BW} by combining all the calibrated datasets in an imaging round with \texttt{WSClean}. The Stokes I images are confusion-limited, with sources at the primary beam's edge appearing elongated due to the beam size changing with frequency over a large bandwidth. This effect is absent in the narrow-band images. In the Stokes I images, we detect numerous background galaxies and diffuse emission from the Milky Way. The Stokes V images are mostly empty, except for the 1\% I$\rightarrow$V residual leakage at the location of the brightest sources, which indicates no detection of Tau Boötis b. The noise levels and resolutions reached in these images are presented in \autoref{tab:noise}. Based on the noise levels, we set 2-sigma upper limits of 64~mJY for Stokes I and 24~mJY for Stokes V.

\begin{table}[!ht]
\centering
\caption{Noise levels and resolutions for the images of the Tau Boötis b field.$\rm ^{(a)}$ }
\label{tab:noise}
\begin{tabular}{cccc}
\hline
\hline
\begin{tabular}[c]{@{}c@{}}Image \\ frequency\end{tabular} & 15~MHz                                                                                      & 24~MHz                                                                                       & 15~MHz-40~MHz                                                                             \\ \hline
$b_{max}$  & 0.34$^\circ$ & 0.26$^\circ$ & 
 0.14$^\circ$ \\
$b_{min}$ & 0.23$^\circ$ &  0.18$^\circ$& 0.09$^\circ$ \\
$PA$ & 132$^\circ$ &  132$^\circ$ & 132$^\circ$ \\
\begin{tabular}[c]{@{}c@{}}Stokes I \\ noise\end{tabular}  & 413~mJy                                                                                     & 304~mJy                                                                                      & 32~mJy                                                                                         \\ 
\begin{tabular}[c]{@{}c@{}}Stokes V\\ noise\end{tabular}   & 91~mJy                                                                                      & 34~mJy                                                                                       & 12~mJy                                                                                        \\ \hline
\end{tabular}
\tablefoot{$\rm ^{(a)}$ $b_{min}$ and $b_{max}$ are the minor and major axis and $PA$ is the position angle.}
\end{table}

\begin{figure*}[!htb]
    \centering
    \begin{subfigure}[t]{\textwidth}
        \centering
        \includegraphics[width = 0.87\linewidth]{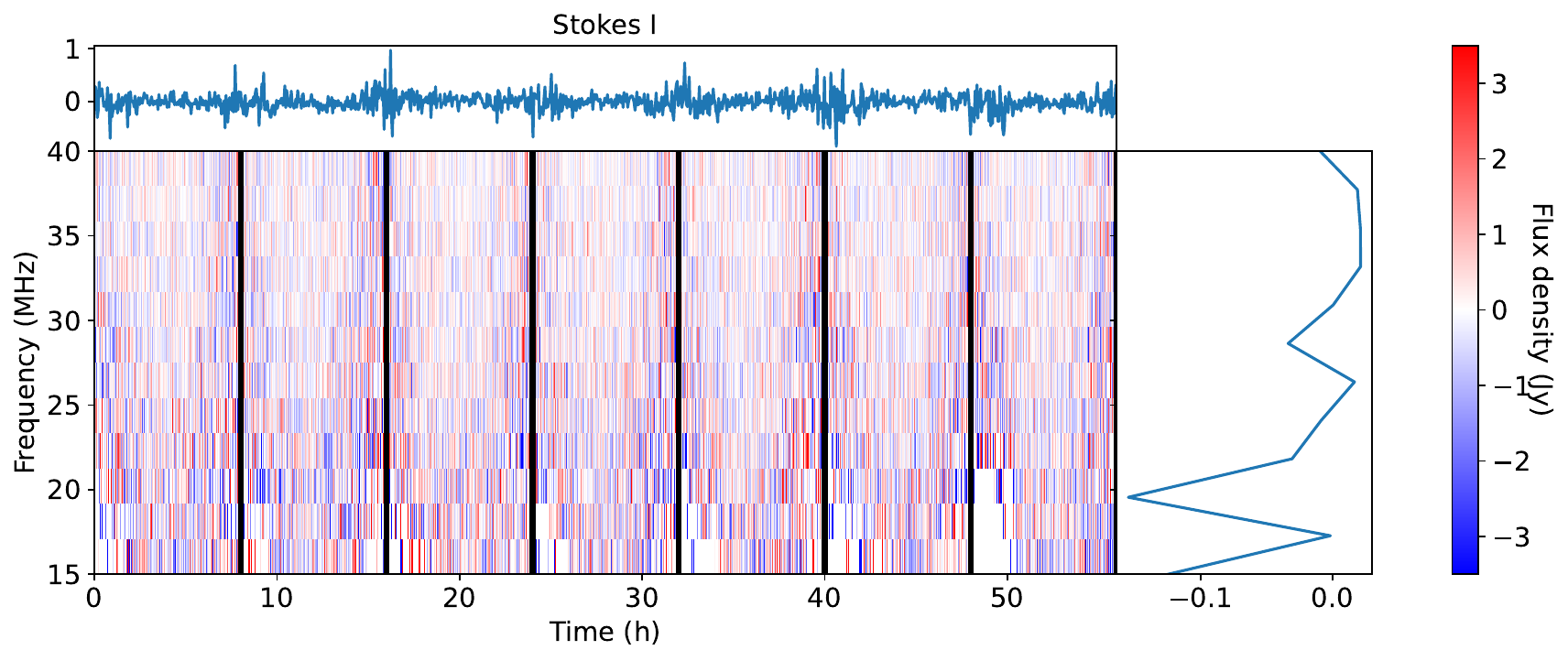}
    \end{subfigure}
    \begin{subfigure}[t]{\textwidth}
        \centering
        \includegraphics[width = 0.87\linewidth]{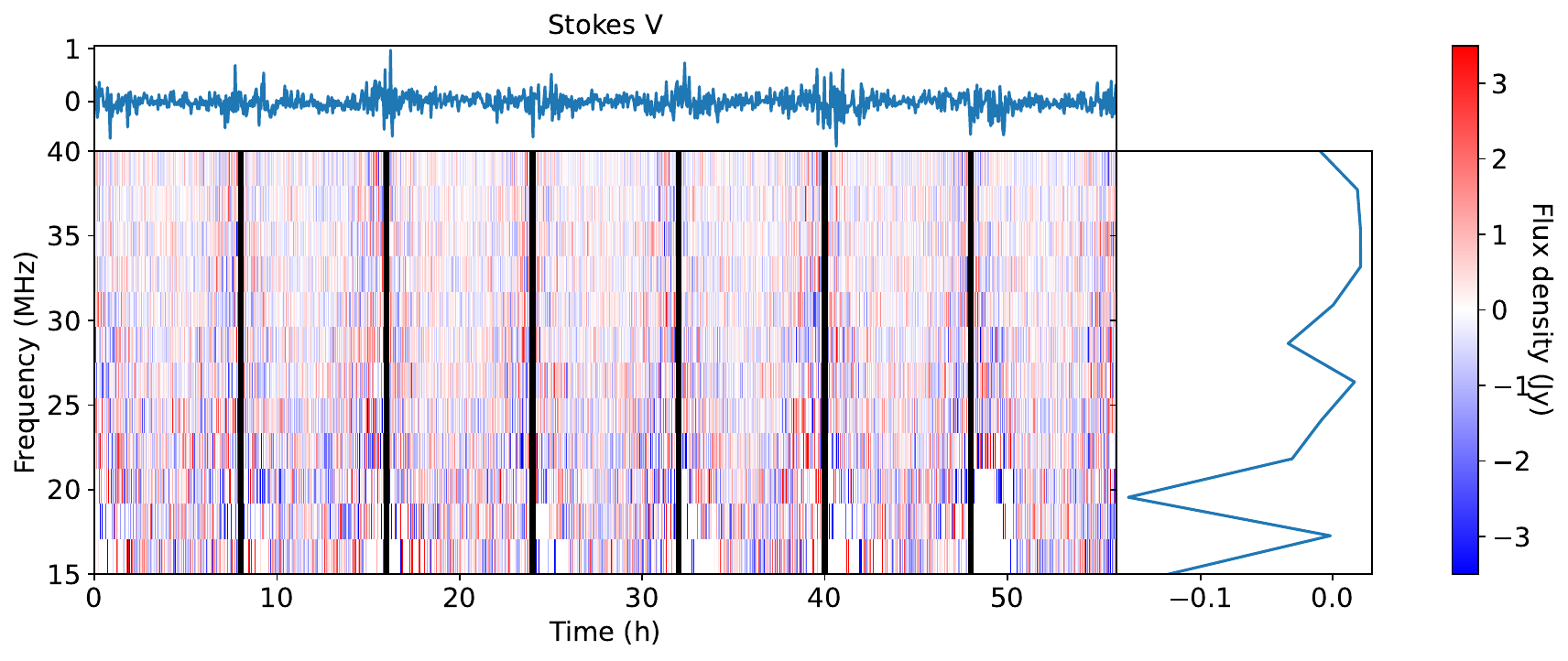}
    \end{subfigure}
    \caption{Dynamic spectrum for different polarisations at the resolution of 1 minute and 1~MHz for all the datasets concatenated. The vertical black lines demarcate the datasets. The noise increases at the beginning and the end of the observation due to the target's low elevation.}
    \label{fig:dynspec}
\end{figure*}

These images represent a 56-hour integration, meaning that short bursts would have been missed due to the long integration time. To search for short-timescale emissions, we created dynamic spectra of the Stokes I and Stokes V emissions at the planet's location (Fig. \ref{fig:dynspec}). For the Stokes I dynamic spectrum, background sources were removed using the model obtained after the last imaging step. The presented spectra have a resolution of 1 minute and 1~MHz, but dynamic spectra with integration times between 4 seconds and 30 minutes were also generated, to match Jupiter's long bursts behaviour. We find no emission above the 3-sigma level for any time and frequency integration. This level is equivalent to 600~mJy for the 1-minute and 1~MHz resolution. The concatenated observations show predominant flagged and un-flagged RFI at the lowest frequencies when the ionosphere reflects obliquely incident interference from distant sources on Earth at the start of each observation.

\begin{figure*}[!htb]
\centering
    \begin{subfigure}[t]{0.45\textwidth}
        \centering
        \includegraphics[width = 1\linewidth]{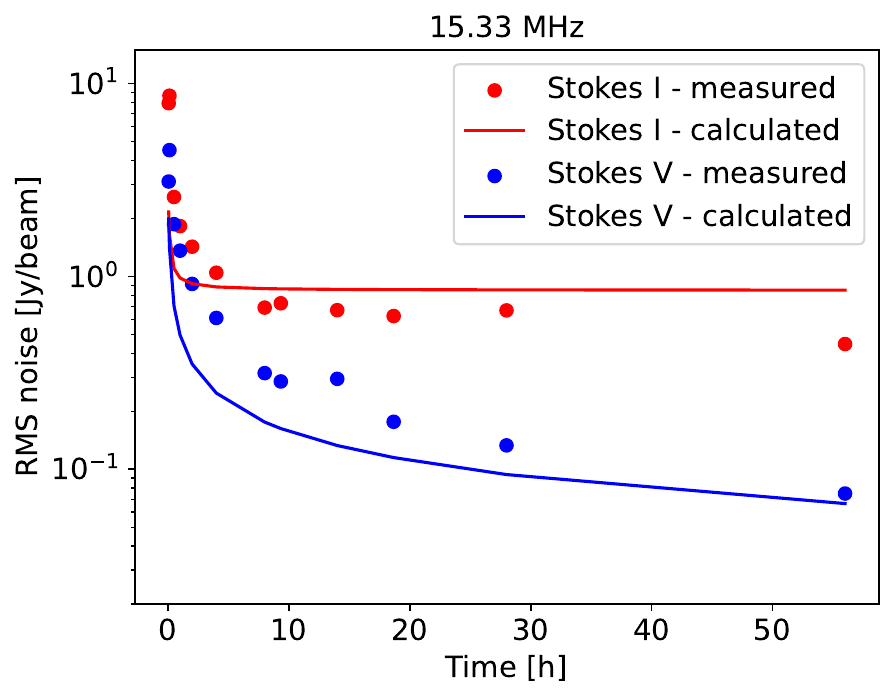}
    \end{subfigure}
    \begin{subfigure}[t]{0.45\textwidth}
        \centering
        \includegraphics[width = 1\linewidth]{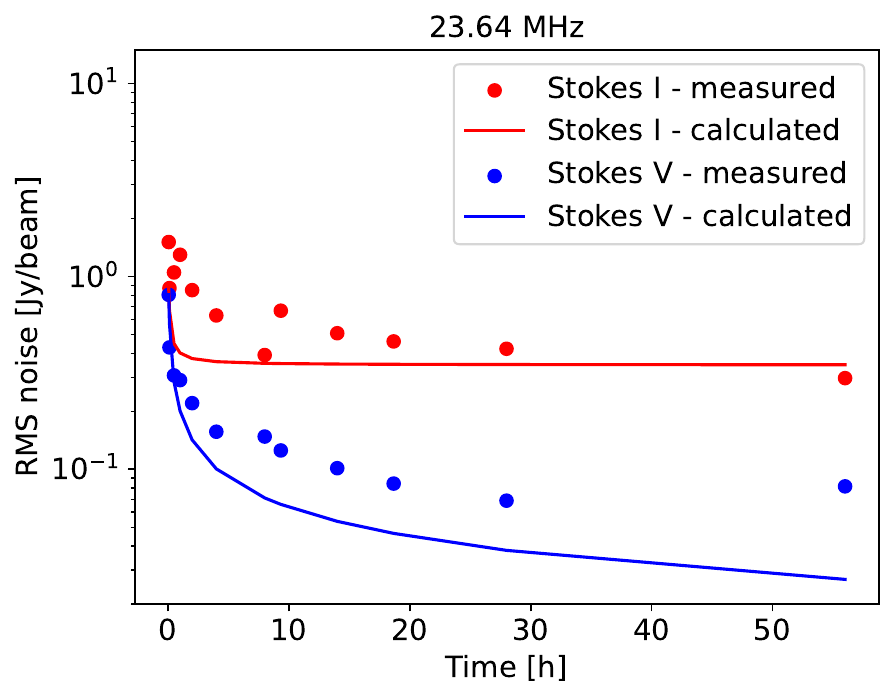}
    \end{subfigure}
    \caption{Measured noise (dots) and predicted noise (line) for the Stokes I (red) and Stokes V (blue) polarisation as a function of time for different frequencies. The bandwidth is 1.2~MHz.}
    \label{fig:noise}
\end{figure*}

To assess the quality of the images, we compared the observational noise with the predicted noise, considering both thermal and confusion noise. The observational noise is the standard deviation of the target field images made with different exposure times. The theoretical noise values are calculated using the LOFAR's system effective flux density and expected source counts (details in \autoref{sec:noise}). We anticipate the noise to be a factor of 2 higher than the theoretical radiometric value due to visibility weighting and the source not always being at high elevation. Additionally, we find that the sidelobes from strong radio sources contribute another factor of 2, as de-mixing with only core stations does not adequately remove A-team sources. Furthermore, flagged data increase the noise by $\sqrt{1-f}$, where $f$ is the fraction of flagged data, so $1-f$ is the fraction of data contributing to the image. After applying all these factors, the results presented in Fig. \ref{fig:noise} show that the predictions align with the measured noise within a factor of 2. At very short time intervals, the noise is higher due to residual RFI, which averages out over longer intervals.

Some measured noise values appear lower than theoretical predictions, likely due to uncertainties in the noise calculation. The lowest frequency at which confusion noise has been reliably calculated is 144 MHz \citep[using LOFAR-][]{mandal2021extremely}. We estimated the confusion noise at 15 MHz by extrapolating with a spectral index of $-0.7$. However, this extrapolation may not accurately represent the noise behaviour at such low frequencies. This limitation reflects the current state of knowledge, and more detailed population studies at low frequencies are required to refine confusion noise estimates.

The thermal noise is also approximated based on data from higher frequencies, with the lowest in situ measurement currently available at 30 MHz \citep{VanHaarlem2013}. It is well established that thermal noise increases rapidly below this frequency, but the exact rate of increase remains uncertain. The noise rise may be less steep than initially assumed and further studies involving antenna performance and system noise at very low frequencies are needed to improve these estimates.

\noindent

\noindent

We present noise plots only for the frequencies corresponding to the images in Fig. \ref{fig:images_freq}. However, the noise in all bands closely matches the predictions. Calculating the predicted confusion noise for the full bandwidth is challenging due to significant beam variations across the band. The noise in the full-bandwidth images does not scale with the square root of the bandwidth, as there is a dramatic increase in sky noise and confusion noise towards lower frequencies. Nonetheless, since the noise for narrow bands matches the predictions, it is reasonable to infer that the noise for the full bandwidth is still close to the theoretical value. Therefore, we did not attempt to approximate it.

\newpage
\section{Discussion}\label{sec:discussion}
Our upper limit for the radio flux density of Tau Boötis b is visualised in Fig. \ref{fig:noise} if the data points are scaled by a factor of 2 on the y-axis to represent a 2-sigma confidence level. Assuming that the radiation is 100\% circularly polarised, as expected from the ECM theory \citep{dulk1985radio}, we establish upper limits for the planet's continuous narrowband and wideband emission in \autoref{tab:limits}. The proximity of a 1~Jy source significantly reduces the likelihood of detecting an unpolarised signal with the current angular resolution. For transient bursts, the required flux increases dramatically, reaching hundreds of mJy or even Jy levels due to the noise scaling with the square root of the integration time.

\begin{table}[!ht]
\centering
\caption{Upper limits of the planet's radio flux for different time resolutions and different polarisations.}
\label{tab:limits}
\begin{tabular}{cccc}
\hline
\hline
\begin{tabular}[c]{@{}c@{}}Integration time\end{tabular} & 56 hours                                                                                      & 8 hours  \\ \hline
\begin{tabular}[c]{@{}c@{}}Unpolarised  radiation \end{tabular}  & 24~mJy                                                                                     & 64~mJy  \\ 
\begin{tabular}[c]{@{}c@{}}Circularly polarised radiation\end{tabular}   & 64~mJy                                                                                      & 170~mJy \\ \hline
\end{tabular}
\end{table}

Our upper limits can be compared with the tentative detection claimed by \citet{turner2021search}, where a flux of 400\,mJy was claimed. Subsequent observations have not detected this emission in either beam-formed \citep{turner2023follow} or interferometric mode (this study) of LOFAR. This non-detection could be attributed to several factors.

Firstly, the visibility of the emission depends on the orientation of the planet's magnetic axis. The magnetic obliquity could have evolved sweeping the ECM beam out of Earth's sightline. However, this is unlikely as gas giant magnetic reversals are expected to occur over timescales of centuries \citep{1986Icar...67...88H} --- much larger than the time span between the claimed detection and our observations.

Another plausible explanation is the variability in stellar activity. Previous studies have suggested stellar activity cycles ranging from 8 months to 1.5 years \citep{catala2007magnetic, donati2008magnetic, fares2009magnetic, fares2013small, mengel2016evolving}, with the most recent findings indicating a period of approximately 120 days \citep{jeffers2018relation}. The star may have exhibited high activity levels in 2016, but since then, its activity may have diminished due to the evolution of its magnetic field or pole reversal. According to radiometric scaling laws \citep{zarka2007plasma}, the radio power from a planet is related to the planetary magnetic field ($B$), stellar magnetospheric standoff distance ($R_s $), and the effective velocity of the stellar wind ($v_{eff}$), as presented in Eq. \ref{eq:radiopower}. If the star's magnetic activity has decreased, this could result in a significant reduction in the stellar wind pressure, leading to weaker interactions between the stellar wind and the planet’s magnetosphere, and consequently a lower radio flux, potentially explaining the non-detection in more recent observations.

\begin{equation}
    P_{radio} \propto v_{eff} B^2 R_s^2
\label{eq:radiopower}
\end{equation}

A final possibility is the potential impact of RFI. In beam-formed data, the emission source cannot be precisely determined, and the tentative detection of \citet{turner2021search} could have been human-generated interference and not a bona fide astrophysical signal.

\begin{figure}[!htb]
    \centering
    \includegraphics[width = 0.98\linewidth]{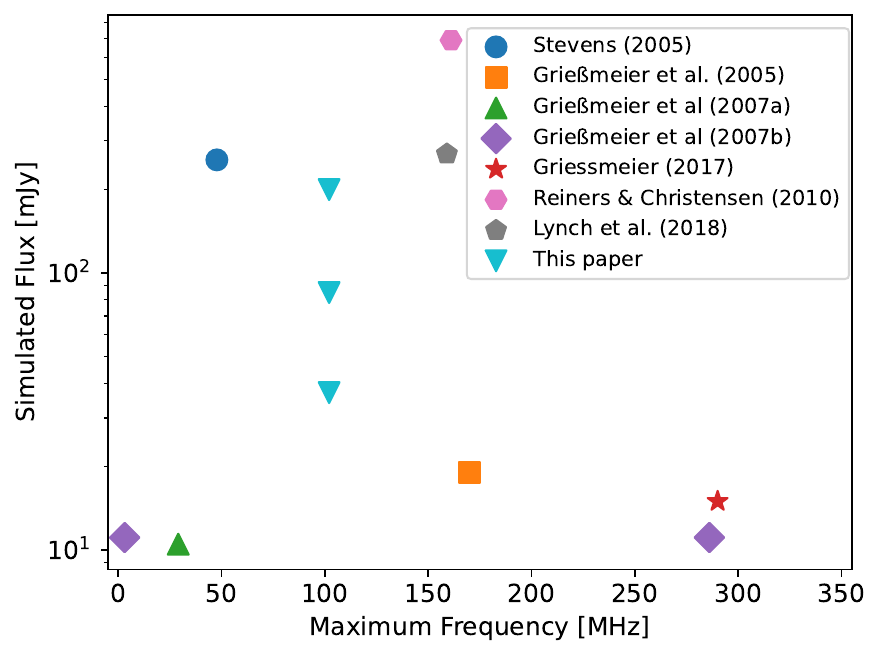}
    \caption{Predicted radio flux densities of Tau Bo\"{o}tis b from different studies. \citet{ griessmeier2007predicting} consider two flux densities, one for the magnetic (higher) and one for the kinetic (lower) incoming flux approximation. We find three flux densities, one for the magnetic approximation (highest), one for the kinetic approximation (middle), and one for when the radio power is saturated (lowest).}
    \label{fig:pred_fluxes}
\end{figure}

Although we could not confirm the tentative detection, our upper limits on Tau Bo\"{o}tis b can be used to constrain models that predict exoplanetary flux densities. These models all use an empirical scaling based on Solar System planetary radio fluxes and assumptions on the host star wind's ram pressure and magnetic pressure. \citet{stevens2005magnetospheric} estimated a flux of 256~mJY below 47.7~MHz, while subsequent studies approximated fluxes between 10~mJY and 300~mJY, all below 20~MHz \citep{griessmeier2005influence, griessmeier2007exoplanetary, griessmeier2007predicting, griessmeier2017search}. Other models have yielded different results. \citet{reiners2010magnetic} predicts a flux of 692.2~mJY below 161~MHz, while \citet{lynch2018detectability} suggests a maximum flux of 269~mJY below 159~MHz. A summary of these studies and their predicted fluxes is presented in Fig. \ref{fig:pred_fluxes}.

We also performed calculations of the predicted flux density broadly following the methodology of \citet{griessmeier2006aspects}. We determined the planet's magnetic field using the approach outlined by \citet{reiners2010magnetic}. The star's coronal temperature, mass loss rate, and magnetic field were calculated using X-ray observations \citep{vidotto2014stellar, wood2021new}. For the magnetic case, we employed a formalism similar to that in \citet{rodriguez2019erosion}. Our predictions indicate a flux of 200~mJy for the kinetic case and 85~mJy for the magnetic case, at a maximum frequency of 102~MHz. The kinetic case refers to the physical scenario when the kinetic energy of the incoming electrons is converted into radio emission, while the magnetic case refers to when the magnetic energy of the electrons is converted to radio emission. These predictions are below the literature values cited above. 

We also considered the possible saturation of the radio power due to limitations in the process converting stellar wind power to planetary radio power, as suggested by \citet{turnpenney2020magnetohydrodynamic}. Using their limiting value of $10^{15}\,{\rm W}$ for the radio power, we estimate a predicted flux density of 37~mJy with a maximum frequency of 102~MHz, which is also marginally inconsistent with our upper limits.

It is crucial to take into consideration the fact that ECM emission is beamed, with the beaming depending on the system's geometry \citep{kavanagh2023hunting}. Our knowledge of the system's geometry is limited. If the initial detection is not real and the planet is not detected again, it may indicate that the planet's emission beam is not oriented towards Earth, making detection impossible. The upper limit discussed above assumes the beam is aligned with the line of sight.

Another possible explanation for the non-detection is that Tau Boötis b's magnetic field may be too weak to generate significant ECM emission in our observing band. The intensity of ECM emission depends on the strength of the planetary magnetic field, and a weaker field would result in lower radio flux and in emission appearing at a lower frequency. This possibility is particularly intriguing because the upper limits reported here are at frequencies comparable to those emitted by Jupiter, providing an opportunity to compare the magnetic properties of Tau Boötis b with those of Solar System planets.

This problem of unknown ECM beaming geometry appears irreducible in targeted observations of known exoplanets. It makes it difficult to use non-detections to critically test flux-prediction models. Instead, an un-targeted sky survey may prove more advantageous, as some very bright radio-emitting planets may have been missed in existing exoplanet searches. In fact, \cite{nichols2011magnetosphere} have suggested that ECM emission powered by exoplanets' rotation (rather than the host stars' wind) is brightest for exoplanets at large orbital distances (exceeding 1\,au). Such planets are more likely to have been missed in existing exoplanet searches that are more sensitive to close-in exoplanets. Our next plan is to search for exoplanets in the LOFAR Decameter Sky Survey \citep[LoDeSS;][]{groeneveld2024characterization}, which uses all the Dutch LOFAR stations to observe between 15 and 30~MHz with 5 hours per pointing, covering the full northern sky.

Additionally, at the end of 2025, LOFAR will be upgraded to LOFAR 2.0 \citep{juerges2021lofar2}. The upgrade will double the amount of low-band antennas. With this enhanced system, we plan to conduct a full-sky survey to search for visible exoplanets on the Dutch sky.

\section{Conclusion} \label{sec:conclusion}

This study aimed to detect decametric radio emissions from the exoplanet Tau Boötis b using LOFAR’s interferometric mode, with observations conducted at frequencies between 15 and 40~MHz. A custom data-processing pipeline was developed to mitigate various challenges, including RFI, ionospheric distortions, and the influence of nearby bright radio sources. Despite observing the planet over eight nights and processing 56 hours of data, no significant Stokes V emission attributable to Tau Boötis b was detected. The non-detection allows us to set upper limits on the planet's radio flux of approximately 30~mJy and challenges previous tentative detections of decametric radio emissions from this exoplanet.

The upper limits derived from these observations are inconsistent with a host of models that predict exoplanet flux densities. Although the non-detection could have been due to Earth not being in the planet's ECM emission beam, our limits prompt a re-examination of the assumptions that go into these models, including the stellar wind density and field strength, the planetary field strength, and the efficiency with which stellar wind energy is converted to planetary radio emission.  

Looking forward, further observational campaigns using more sensitive instruments, such as the upcoming LOFAR 2.0 upgrade, will allow for deeper imaging below 40~MHz, where exoplanetary signals are anticipated. Expanding these searches to include large-scale sky surveys will also help us detect rotationally powered radio emission from exoplanets with longer periods that were missed in traditional exoplanet searches.

\begin{acknowledgements}
      CMC and HKV acknowledge funding from the European Research Council via the starting grant `STORMCHASER' (grant number 101042416). CMC thanks Dr. Joe Callingham for discussions. This work made use of the python packages \texttt{matplotlib} \citep{mpl} to generate the figures and \texttt{numpy} \citep{np} for computations. 
\end{acknowledgements}

%
%
\bibliographystyle{aa}
\bibliography{aa52868-24corr} 

\begin{appendix}

\onecolumn
\section{Calibrator phase solutions} \label{sec:phasesol}

\begin{figure}[hb]
    \centering
    \includegraphics[width=0.9\linewidth]{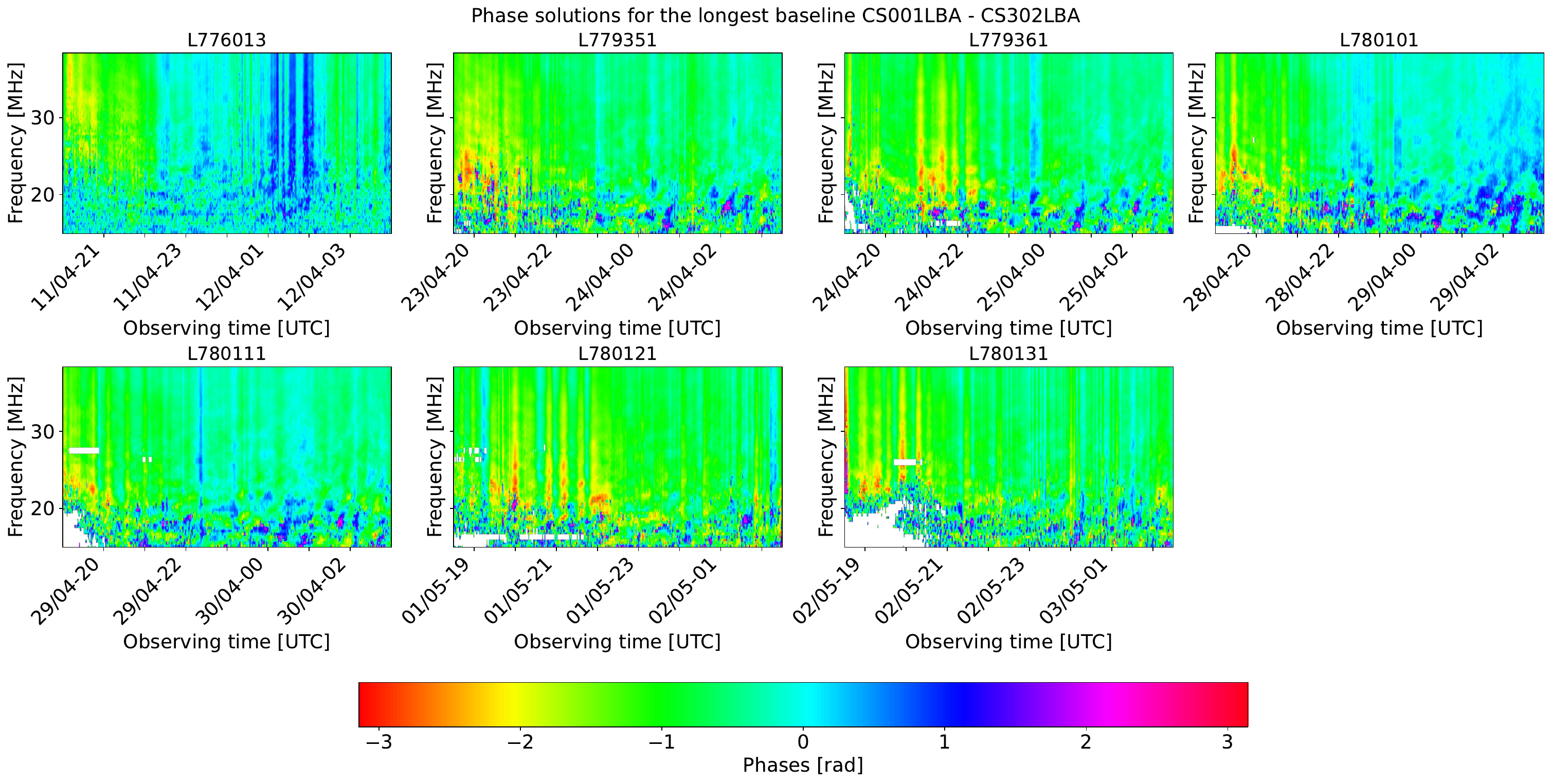}
    \caption{Phase solutions of the calibrator for the longest baseline.}
    \label{fig:phasecal}
\end{figure}

The calibrator phase solutions for the longest baseline provide a representation of the ionospheric conditions at the time of the observation. In our case, the longest baseline is CS001-CS302 and the phase solutions for all observations are presented in Fig. \ref{fig:phasecal}. The phases are changing slowly and there are smooth transitions between the frequency and time, with no phase wraps. All these points towards highly constant (and therefore calm) ionospheric conditions.

\FloatBarrier

\section{Calibrator images} \label{sec:imagesCal}

\begin{figure*}[h!t]
    \centering
    \begin{subfigure}[t]{0.4\textwidth}
        \centering
        \includegraphics[width = 1\linewidth]{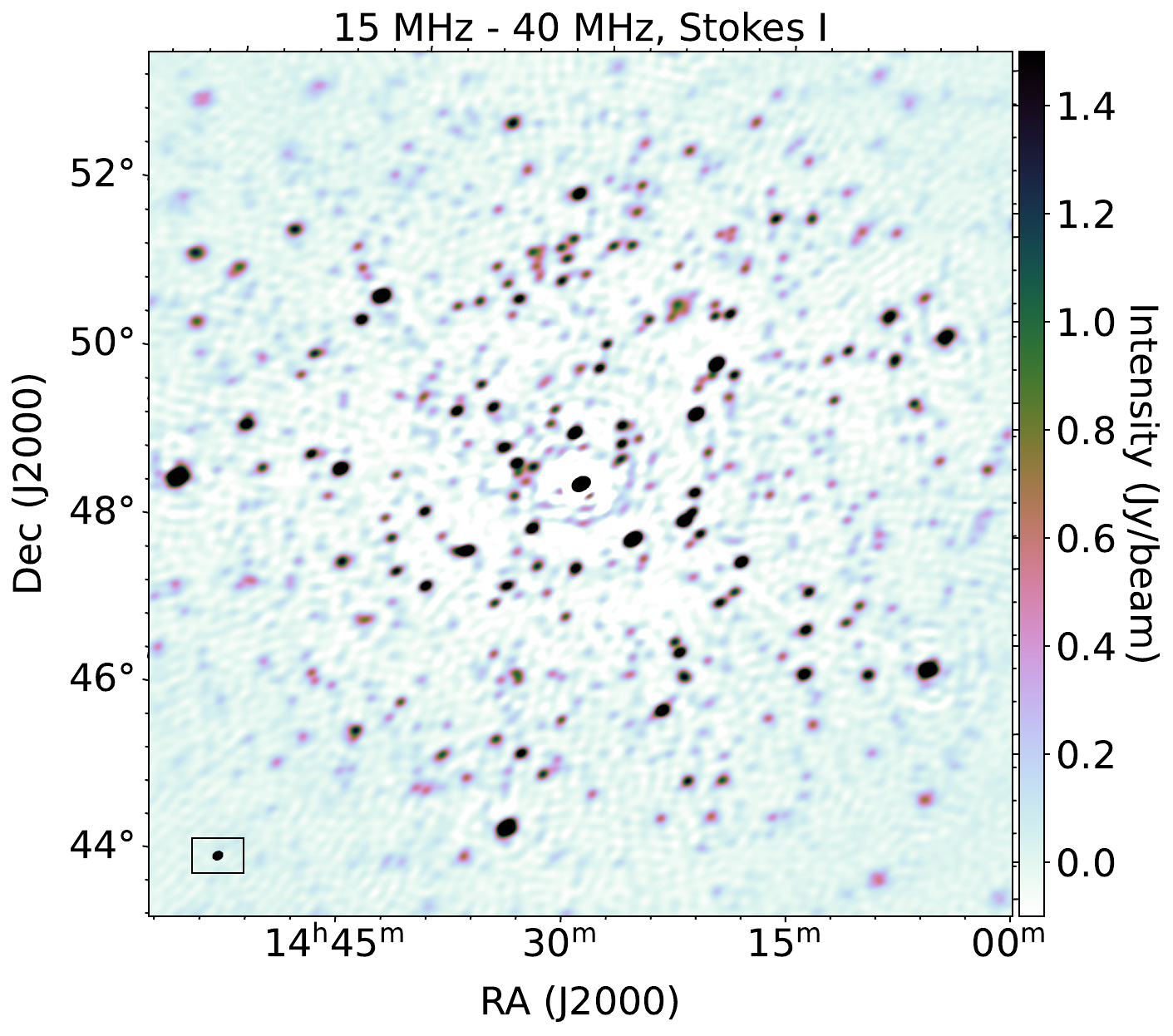}
    \end{subfigure}
    \begin{subfigure}[t]{0.4\textwidth}
        \centering
        \includegraphics[width = 1\linewidth]{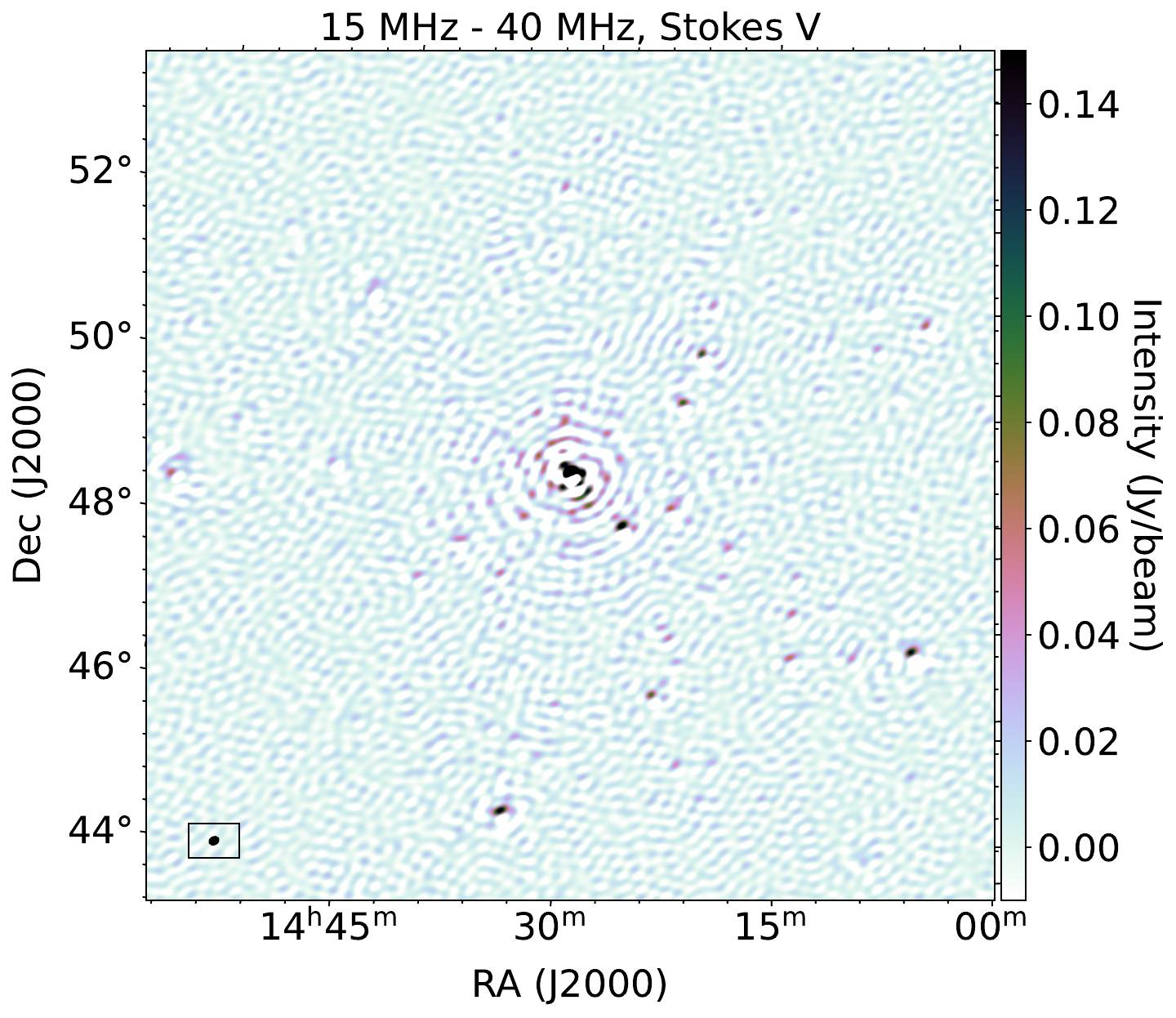}
    \end{subfigure}
    \caption{Calibrator images in different polarisations for an integration time of 56 hours and a bandwidth of 35~MHz. The beam shape is in the lower-left corner. The leaked flux from Stokes I to Stokes V is 2 Jy (3\%).}
    \label{fig:cal_BW}
\end{figure*}

\begin{figure*}[h!t]
    \centering
    \begin{subfigure}[t]{0.4\textwidth}
        \centering
        \includegraphics[width = 1\linewidth]{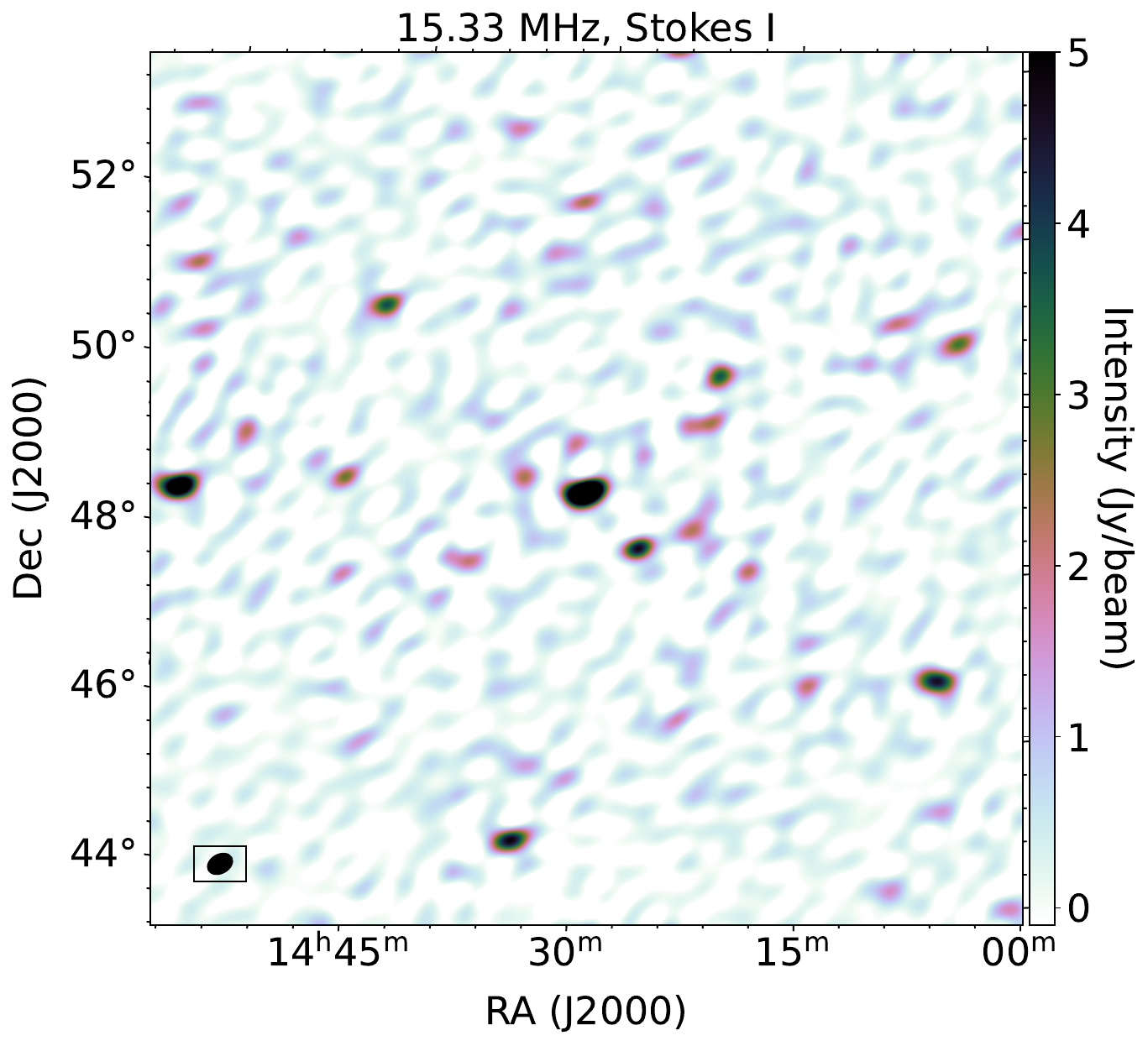}
    \end{subfigure}%
    ~ 
    \begin{subfigure}[t]{0.4\textwidth}
        \centering
        \includegraphics[width = 1\linewidth]{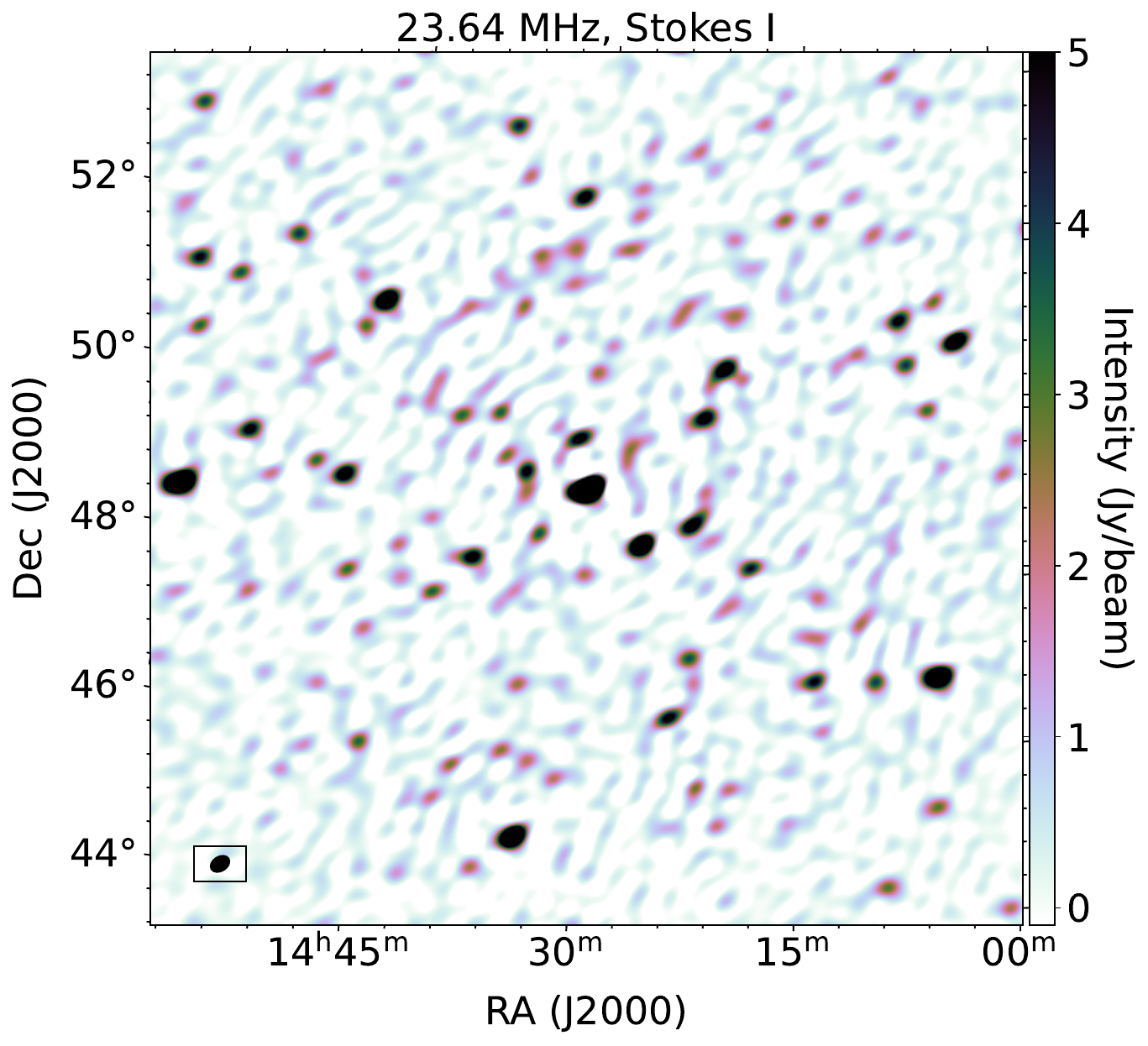}
    \end{subfigure}

    \begin{subfigure}[t]{0.4\textwidth}
        \centering
        \includegraphics[width = 1\linewidth]{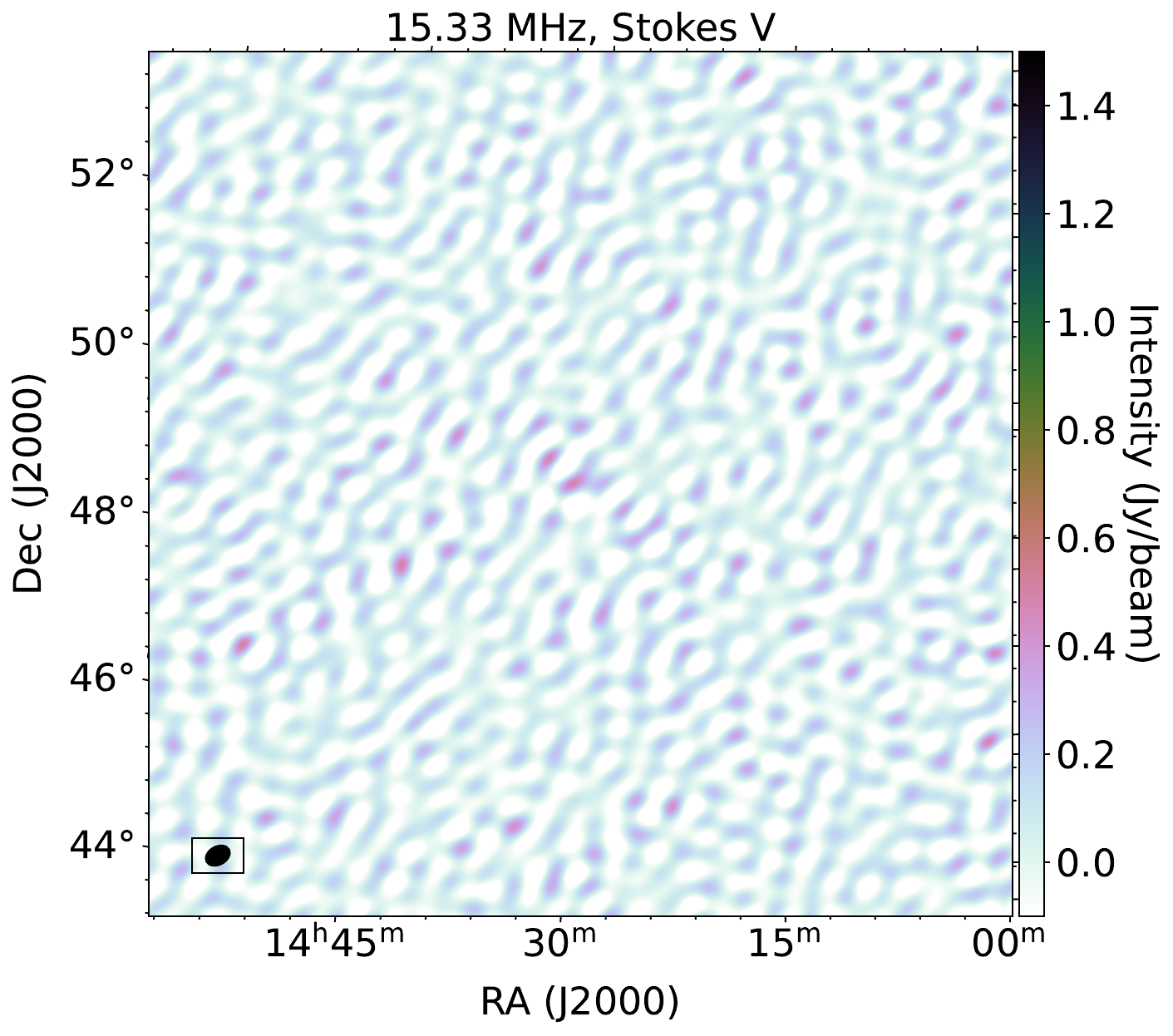}
    \end{subfigure}%
    ~ 
    \begin{subfigure}[t]{0.4\textwidth}
        \centering
        \includegraphics[width = 1\linewidth]{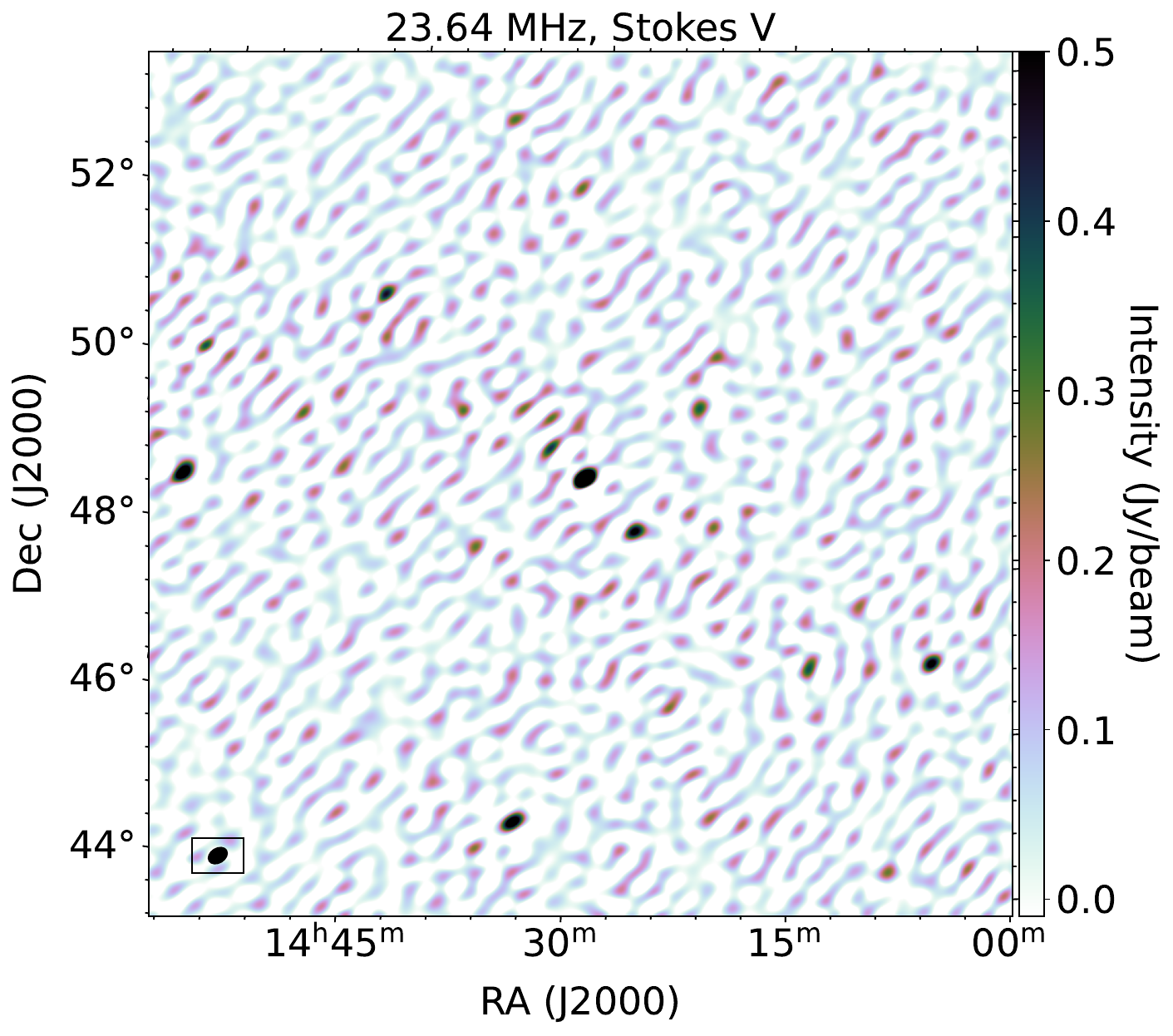}
    \end{subfigure}
    \caption{Calibrator images at 15~MHz and 24~MHz in different polarisations for an integration time of 56 hours and a bandwidth of 1.2~MHz. The beam shape is in the lower-left corner. The leaked flux from Stokes I to Stokes V is below the noise at 15~MHz and 2 Jy (5\%) at 24~MHz.}
    \label{fig:cal_freq}
\end{figure*}

We present the images of the calibrator similarly to the target images presented in the main text. The full bandwidth images are presented in Fig. \ref{fig:cal_BW}. The images at 1.2~MHz in Stokes I and Stokes V at 15~MHz and 24~MHz are presented in Fig. \ref{fig:cal_freq}. These images are not corrected for the leakage from Stoke I to Stokes V, so the Stokes V images present 10\% of signal leakage. The noise levels and resolutions of these images are presented in \autoref{tab:noise_cal}.

\begin{table}[!ht]
\centering
\caption{Noise levels and resolutions for the calibrator field. $\rm ^{(a)}$}
\label{tab:noise_cal}
\begin{tabular}{cccc}
\hline
\hline
\begin{tabular}[c]{@{}c@{}}Image \\ frequency\end{tabular} & 15~MHz                                                                                      & 24~MHz                                                                                       & 15~MHz-40~MHz                                                                             \\ \hline
$b_{max}$ & 0.31$^\circ$ & 0.24$^\circ$ & 0.12$^\circ$ \\ 
$b_{min}$ & 0.21$^\circ$ &0.17$^\circ$  &  0.09$^\circ$\\
$PA$ & 119$^\circ$ & 119$^\circ$ &  119$^\circ$\\
\begin{tabular}[c]{@{}c@{}}Stokes I \\ noise\end{tabular}  & 395~mJy                                                                                     & 273~mJy                                                                                      & 26~mJy                                                                                         \\ 
\begin{tabular}[c]{@{}c@{}}Stokes V\\ noise\end{tabular}   & 97~mJy                                                                                      & 45~mJy                                                                                       & 10~mJy                                                                                        \\ \hline
\end{tabular}
\tablefoot{$\rm ^{(a)}$ $b_{min}$ and $b_{max}$ are the minor and major axis, and $PA$ is the position angle.}
\end{table}
\FloatBarrier

\section{Self-calibration phase solutions} \label{sec:phasetarget}

\begin{figure*}[h!t]
    \centering
    \includegraphics[width=1\linewidth]{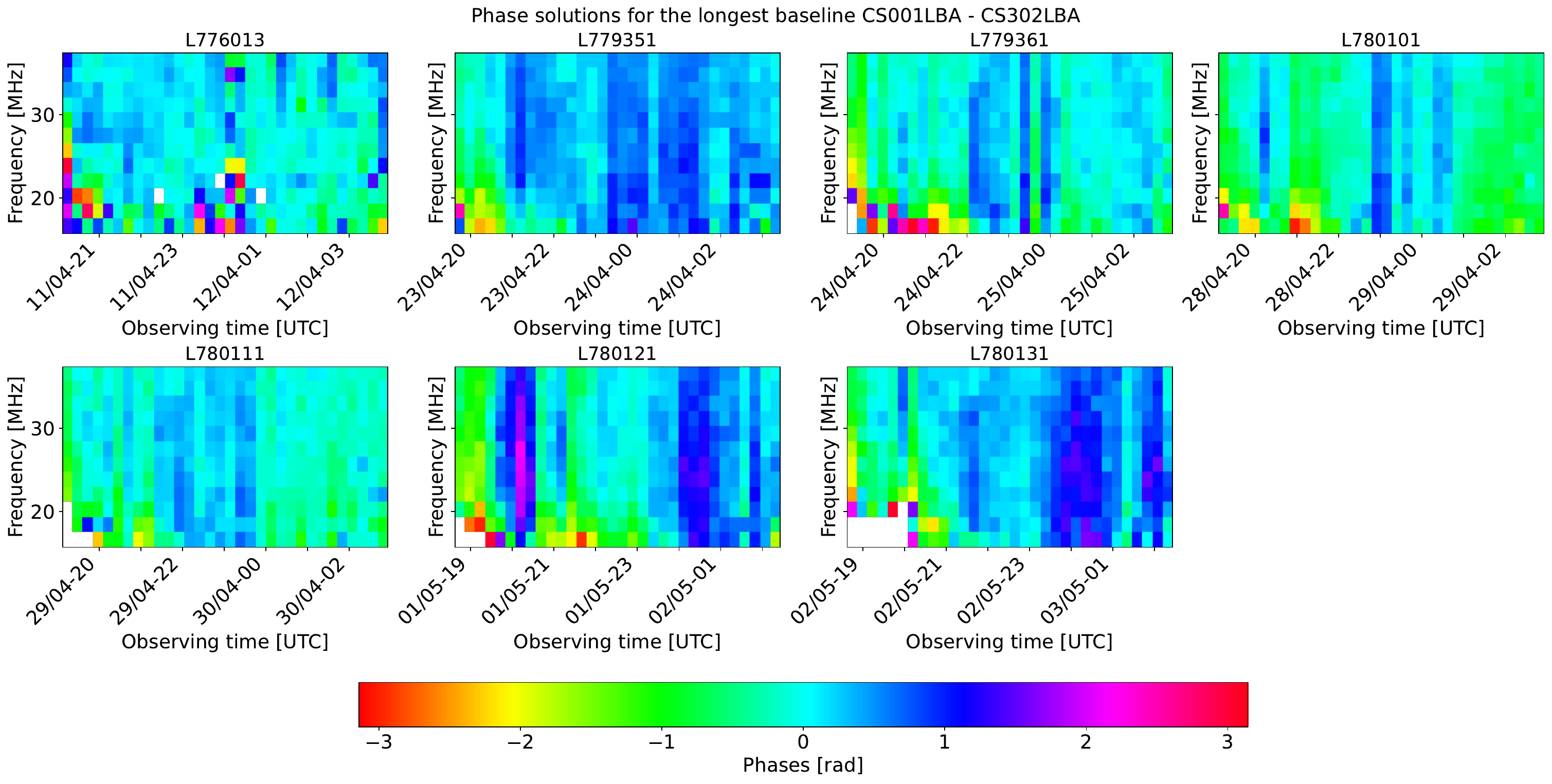}
    \caption{Phase solutions of the target self-calibration step for the longest baseline.}
    \label{fig:phasetarget}
\end{figure*}

The target phase solutions for the longest baseline provide a representation of the ionospheric conditions at the time of the observation. In our case, the longest baseline is CS001-CS302 and the phase solutions for all observations are presented in Fig. \ref{fig:phasetarget}. The phases are changing slowly and there are smooth transitions between the frequency and time, with no phase wraps. All these points towards highly constant (and therefore calm) ionospheric conditions.

\FloatBarrier

\section{Noise calculation} \label{sec:noise}
The total noise ($\sigma$) in radio observations is a combination of the thermal ($\sigma_t$) and confusion noise ($\sigma_c$). They are added in quadrature because they are independent of each other (Eq. \ref{eq:noise-total}).

\begin{equation}
    \sigma = \sqrt{\sigma_t^2 + \sigma_c^2}
    \label{eq:noise-total}
\end{equation}

Confusion noise is a type of background noise that arises when multiple sources are indistinguishable from one another. This often occurs where the instrument's resolution is insufficient to separate closely spaced sources. As a result, the signals from these sources blend, creating a `confused' signal that can obscure or distort the measurement of individual sources. Confusion limit occurs when the source density is roughly one source per 13 beam areas, as shown in Eq. \ref{eq:limit}, where $\Omega$ is the beam size of the observation and $N(\sigma_c)$ is the number of sources per steradian at the confusion limit \citep{cohen2004estimates}. The beam solid angle can be calculated using Eq. \ref{eq:solid-angle}, where we assumed the beam to be a Gaussian with the two main axes $\theta_1$ and $\theta_2$.

\begin{equation}
    12.9 \ \Omega \ N(\sigma_c) = 1 \ \implies N(\sigma_c) = \frac{1}{12.9 \ \Omega}
    \label{eq:limit}
\end{equation}

\begin{equation}
    \Omega = \theta_1  \ \theta_2
    \label{eq:solid-angle}
\end{equation}

The number of sources per steradian with a flux density larger than $s$ ($N(s)$) was previously parameterised in the  Very Large Array \citep[VLA; ][]{thompson1980very} Sky Survey at 74~MHz \citep{cohen2007vla,lane2012vlss} and it presented in Eq. \ref{eq:N-s}, where $\lambda$ is the observation wavelength, $\lambda_0$, a reference wavelength, $\alpha$, the intrinsic slope of the underlying source distribution, $\beta$, the mean spectral index of a source, and A, an unknown constant.
\begin{equation}
    N(s) = A \ s^\alpha \left( \frac{\lambda}{\lambda_0} \right)^{\beta \alpha}
    \label{eq:N-s}
\end{equation}

We write Eq. \ref{eq:N-s} for the confusion noise ($N(\sigma_c)$) and a known measurement \citep[$N(s_1)$ -][]{mandal2021extremely} in Eq. \ref{eq:two-eq}. Then we calculate their ratio and extract the confusion noise using Eq. \ref{eq:confusion-noise}, where $N(\sigma_c)$ is calculated in Eq. \ref{eq:limit}.
\begin{equation}
    \begin{split}
        & N(\sigma_c) = A \ \sigma_c^\alpha \left( \frac{\lambda_c}{\lambda_0} \right)^{\beta \alpha} \\
        & N(s_1) = A \ s_1^\alpha \left( \frac{\lambda_1}{\lambda_0} \right)^{\beta \alpha}
    \end{split}
    \label{eq:two-eq}
\end{equation}

\begin{equation}
    \frac{N(s_1)}{N(\sigma_c)} = \left( \frac{s_1}{\sigma_c} \right)^\alpha \left( \frac{\lambda_1}{\lambda_c} \right) ^{\beta \alpha}  \ \implies \sigma_c = s_1 \left( \frac{\lambda_1}{\lambda_c} \right) ^{\beta}  \left( \frac{N(\sigma_c)}{N(s_1)} \right) ^{1/\alpha}
    \label{eq:confusion-noise}
\end{equation}

The system's effective flux density is generally calculated using Eq. \ref{eq:SEFD}, where $k$ is the Boltzmann's constant, $\eta$, the system efficiency factor ($\sim$ 1.0), $A_{eff}$, the total collecting area and $T_{eff}$, the system's noise temperature. For LOFAR, the system's noise temperature is dominated by galactic radiation, and it has a strong wavelength dependence shown in Eq. \ref{eq:T-gal}, where $T_0 = 60 \pm 20 \ K$ for galactic latitudes between 10$^\circ$ and 90$^\circ$. The collecting area is calculated using Eq. \ref{eq:area}, where d is the distance to the nearest dipole and $N$ is the number of stations used in the observation.

\begin{equation}
    S_{sys} = \frac{2 \eta k}{A_{eff}} T_{sys}
    \label{eq:SEFD}
\end{equation}

\begin{equation}
    T_{sys} = T_0 \lambda^{2.55}
    \label{eq:T-gal}
\end{equation}

\begin{equation}
    A_{eff} = N \ min \left\{ \frac{\lambda^2}{3}, \frac{\pi d^2 }{4} \right\}
    \label{eq:area}
\end{equation}

The thermal noise is calculated from the system's effective flux density using Eq. \ref{eq:thermal-noise}, where $N$ is the number of stations, $\Delta\nu$, the bandwidth, $\Delta t$, the integration time, and the factor of 2 appears because each antenna has two dipoles perpendicular to each other.

\begin{equation}
    \sigma_t = \frac{S_{sys}}{\sqrt{N (N-1)2\Delta\nu \Delta t}}
    \label{eq:thermal-noise}
\end{equation}

\end{appendix}
\end{document}